\documentclass[11pt]{article}

\usepackage{amssymb}
\usepackage{citesort}
\usepackage[cp1250]{inputenc}
\usepackage[dvips]{graphicx}
\newcommand{\bra}[1]{\langle{\rm #1}|}
\newcommand{\ket}[1]{|{\rm #1}\rangle}

\renewcommand{\theequation}{\arabic{section}.\arabic{equation}}

\begin{document}
\begin{titlepage}
\thispagestyle{empty}

\begin{flushright}
                hep-th/0504204\\
                TPJU -- 5/2005\\
                0104/2005 IFT UWr
\end{flushright}
\bigskip

\begin{center}

{\LARGE\bf\sf Classical geometry from the quantum Liouville theory} \\

\end{center}
\bigskip

\begin{center}
    {\large\bf\sf
    Leszek Hadasz}\footnote{e-mail: hadasz@th.if.uj.edu.pl} \\
\vskip 1mm
    M. Smoluchowski Institute of Physics, \\
    Jagellonian University,
    Reymonta 4, 30-059 Krak\'ow, Poland \\
\vskip 5mm
    {\large\bf\sf
    Zbigniew Jask\'{o}lski}\footnote{e-mail: jask@ift.uni.wroc.pl
}\\
{\large\bf\sf
    Marcin Pi\c{a}tek}\footnote{e-mail: piatek@ift.uni.wroc.pl
}\\
\vskip 1mm
   Institute of Theoretical Physics\\
University of Wroc{\l}aw\\
 pl. M. Borna, 950-204 Wroc{\l}aw
 Poland
\end{center}

\vskip .5cm

\begin{abstract}
Zamolodchikov's recursion relations are used to analyze the
existence and approximations to the classical conformal block in
the case of four parabolic weights. Strong numerical evidence is
found that the saddle point momenta arising in the classical limit
of the DOZZ quantum Liouville theory are simply related to the
geodesic length functions of the hyperbolic geometry on the
4-punctured Riemann sphere. Such relation provides new powerful
methods for both numerical and analytical calculations of these
functions. The consistency conditions for the factorization of the
4-point classical Liouville action in different channels are
numerically verified. The factorization yields efficient numerical
methods to calculate the 4-point  classical action and, by the
Polyakov conjecture, the accessory parameters of the Fuchsian
uniformization of the 4-punctured sphere.

\end{abstract}

\vspace{\fill}

\vspace*{5mm}
\end{titlepage}
\newpage
\vskip 1cm
\section{Introduction}
\setcounter{equation}{0}

A few years ago a considerable progress in the Liouville theory has been achieved \cite{Teschner:2001rv}.
The solution to the quantum theory based on the structure constants
proposed by Otto and Dorn~\cite{Dorn:1994xn}
and by A. and Al. Zamolodchikov \cite{Zamolodchikov:1995aa}
was completed by  Ponsot and Teschner~\cite{Ponsot:1999uf,Ponsot:2000mt,Ponsot:2001ng}
and by Teschner \cite{Teschner:2001rv,Teschner:2003en}.
Along with the techniques of calculating conformal blocks developed
by Al. Zamolodchikov  \cite{Zamolodchikov:ie,Zam0,Zam} the DOZZ theory provides explicit formulae for
quantum correlators.

On the other hand there exists so called geometric approach
originally proposed by Polyakov \cite{Polyakov82} and further
developed by Takhtajan
\cite{Takhtajan:yk,Takhtajan:1994vt,Takhtajan:1993vt,Takhtajan:zi}
(see also
\cite{Matone:1993tj,Matone:1993nf,Matone:1994pz,Bertoldi:2004yk}).
In contrast to the operator formulation of the DOZZ theory the
correlators of primary fields with elliptic and parabolic weights
are expressed in terms path integral
 over conformal class of Riemannian metrics
with prescribed  singularities at the punctures.
The underlying structure of this formulation
is the classical hyperbolic geometry of the Riemann surface.

Although the  relation between these formulations is not yet
completely understood
\cite{Takhtajan:1995fd,Menotti:2004xz,Menotti:2004uq} it is
commonly believed that the quasiclassical limit of the DOZZ theory
exists and is correctly described  by the classical Liouville
action of the geometric approach. This is for instance justified
by explicit calculation of the classical limit of the DOZZ
structure constants and the classical Liouville action for the
Riemann sphere with three punctures
\cite{Zamolodchikov:1995aa,Hadasz:2003he}.

Some  of the predictions derived from the path integral
representation of the geometric approach
can be rigorously proved and lead to deep geometrical results. This can be seen as
 an additional support for the correctness of the
picture  the geometric formulation
provides for the semiclassical limit of the DOZZ theory.
One of the results of this type
is the so called Polyakov conjecture
obtained  as a classical limit of the Ward identity
\cite{ZoTa1,ZoTa2,Zograf,Cantini:2001wr,Takhtajan:2001uj,Hadasz:2003kp}.
It states that the classical Liouville action is a generating
function for the  accessory parameters of the Fuchsian
uniformization of the punctured sphere
yielding an essentially new insight into this classical long standing problem.
Its usefulness for solving the uniformization
is however restricted by our ability to calculate the classical Liouville action for
more than three singularities.

The existence of the semiclassical limit of the Liouville correlation function
with the projection on one intermediate conformal family implies a
semiclassical
 limit of the BPZ quantum conformal block \cite{Belavin:1984vu} with heavy weights $\Delta=Q^2\delta$,
$\Delta_i=Q^2\delta_i,$ with $\delta,\delta_i = {\cal O}(1)$ in the following form:
\begin{equation}
\label{defccb}
{\cal F}_{\!1+6Q^2,\Delta}
\!\left[_{\Delta_{4}\;\Delta_{1}}^{\Delta_{3}\;\Delta_{2}}\right]
\!(x)
\; \sim \;
\exp \left\{
Q^2\,f_{\delta}
\!\left[_{\delta_{4}\;\delta_{1}}^{\delta_{3}\;\delta_{2}}\right]
\!(x)
\right\}.
\end{equation}
The function $\,f_{\delta}
\!\left[_{\delta_{4}\;\delta_{1}}^{\delta_{3}\;\delta_{2}}\right](x)$ is  called the
classical conformal block \cite{Zamolodchikov:1995aa} or
(a bit confusing) the ``classical action'' \cite{Zam0,Zam}.

The existence of the semiclassical limit (\ref{defccb}) was first
postulated in \cite{Zam0,Zam} where it was pointed out that the
classical block is related to a certain monodromy problem
of a null vector decoupling equation in a similar
way the classical Liouville action is related to the
Fuchsian uniformization. This relation was further used to
derive the $\Delta\to \infty$ limit of the conformal block
and its expansion in powers of the $q$ variable.

The 4-point function of the DOZZ theory can be  defined as an
integral of $s$-channel conformal blocks and the DOZZ couplings
over
 the continuous spectrum of the theory.
In the semiclassical limit the integrand can be expressed in terms
of the 3-point classical Liouville action and the classical block,
and the integral itself is dominated by the saddle point
$\Delta_s=Q^2\delta_s(x)$. One thus gets the factorization
\cite{Zamolodchikov:1995aa}
\begin{eqnarray}
\label{clasfact}
&&\hspace{-30pt} S^{\rm (cl)}(\delta_4,\delta_3,\delta_2,\delta_1;x) =\\
&& S^{\rm (cl)}(\delta_4,\delta_3,\delta_s(x))
+
S^{\rm (cl)}(\delta_s(x),\delta_2,\delta_1)
-\,f_{\delta_s(x)}
\!\left[_{\delta_{4}\;\delta_{1}}^{\delta_{3}\;\delta_{2}}\right](x)
-\bar f_{\delta_s(x)}
\!\left[_{\delta_{4}\;\delta_{1}}^{\delta_{3}\;\delta_{2}}\right](\bar x),
\nonumber
\end{eqnarray}
where $S^{\rm (cl)}(\delta_4,\delta_3,\delta_2,\delta_1;x)$ is the classical
action for the weights $\delta_1,\delta_2,\delta_3, \delta_4$ located
at $0,x,1$ and $\infty$, respectively, and $S^{\rm (cl)}(\delta_3,\delta_2,\delta_1)$
is the classical Liouville action for the weights $\delta_1,\delta_2,\delta_3$
at the locations $0,1,\infty$.
Since the semiclassical limit should be independent of the choice of the channel in
the representation of the DOZZ 4-point function one gets the consistency conditions
\begin{eqnarray}
\nonumber
&& \hspace{-100pt} S^{\rm (cl)}(\delta_4,\delta_3,\delta_s(x))
+
S^{\rm (cl)}(\delta_s(z),\delta_2,\delta_1)
-\,f_{\delta_s(x)}
\!\left[_{\delta_{4}\;\delta_{1}}^{\delta_{3}\;\delta_{2}}\right](x)
-\bar f_{\delta_s(x)}
\!\left[_{\delta_{4}\;\delta_{1}}^{\delta_{3}\;\delta_{2}}\right](\bar x)
\\[10pt]
\label{clasboot}
&=&
S^{\rm (cl)}(\delta_4,\delta_1,\delta_t(x))
+
S^{\rm (cl)}(\delta_t(x),\delta_2,\delta_3)\\
\nonumber
&&
\hspace{70pt}-\,f_{\delta_t(x)}
\!\left[_{\delta_{4}\;\delta_{3}}^{\delta_{1}\;\delta_{2}}\right](1-x)
-\bar f_{\delta_t(x)}
\!\left[_{\delta_{4}\;\delta_{3}}^{\delta_{1}\;\delta_{2}}\right](1-\bar x)
\\[10pt]
\nonumber
&=&  2\delta_2 \log x\bar x + S^{\rm (cl)}(\delta_1,\delta_3,\delta_u(x))
+
S^{\rm (cl)}(\delta_u(x),\delta_2,\delta_4)\\
\nonumber
&&
\hspace{70pt}-\,f_{\delta_u(x)}
\!\left[_{\delta_{1}\;\delta_{4}}^{\delta_{3}\;\delta_{2}}\right]\left({1\over x}\right)
-\bar f_{\delta_u(x)}
\!\left[_{\delta_{1}\;\delta_{4}}^{\delta_{3}\;\delta_{2}}\right]\left({1\over \bar x}\right),
\end{eqnarray}
where the saddle weights $\delta_t(x), \delta_u(x)$ in the $t$- and $u$-channel  are simply related to
the $s$-channel saddle point:
$$
\delta_t(x) = \delta_s(1-x),\;\;\;\;\;\;\delta_u(x) = \delta_s\left({1\over x}\right).
$$
As the conditions (\ref{clasboot}) follow from the semiclassical limit
of the  bootstrap equations of the quantum DOZZ theory
we shall
call them the {\it classical bootstrap equations}.

The question arises what is the geometric interpretation of the saddle point conformal weight
$\delta_s(x)$. Let us recall that the classical solution describes a unique hyperbolic geometry
with singularities at the locations of conformal weights. For
 elliptic, parabolic and hyperbolic weights one gets
conical singularities, punctures and holes with geodesic
boundaries, respectively \cite{sei,Hadasz:2003he,Hadasz:2003kp}.
In the latter case the (classical) conformal weight $\delta$ is
related to the length $\ell$ of the corresponding geodesic by
\begin{equation}
\label{glength}
\delta =\frac{1}{4}+
\frac{\mu}{4}\left(\frac{\ell}{2\pi}\right)^2
\end{equation}
where the scale of the classical configuration is set by the condition $R=-{\mu\over 2}$
imposed on the constant scalar curvature $R$.

Let us note that the relation (\ref{glength}) appears also in the
context of the quantization of the Teichm\"uller space
(\cite{Verlinde:1989ua}, \cite{Teschner:2003em} and references
therein). With this relation assumed one can show that the Hilbert
space arising  in the quantization of the Teichm\"uller space of a
Riemann surface and the space of conformal blocks of the Liouville
theory on this surface are isomorphic as representations of the
mapping class group.

In the case of 4 singularities at the standard locations
$0,x,1,\infty$ there are  three closed ge\-o\-de\-sics $\Gamma_s,
\Gamma_t, \Gamma_u$ separating the singular points into pairs
$(x,0|1,\infty),$ $(x,1|0,\infty)$ and $(x,\infty|0,1)$
respectively. Since the spectrum of DOZZ theory is hyperbolic the
singularities corresponding to the saddle point weights
$\delta_i(x)$ are geodesic holes. One may expect that these
weights are related to the lengths $\ell_i$ of the closed
geodesics $\Gamma_i$ in corresponding channels:
\begin{equation}
\label{glengthII} \delta_i(x) =\frac{1}{4}+
\frac{\mu}{4}\left(\frac{\ell_i(x)}{2\pi}\right)^2,\;\;\;i=s,t,u.
\end{equation}
If the formula above proved true
it would provide a new powerful tool for calculating the geodesic length functions which
play an important role in analyzing the structure of the moduli space of Riemann surfaces
(\cite{Wolf}, and references therein).
It would also pave a way for an explicit uniformization of at least 4-punctured sphere.

Our aim in the present paper is a numerical verification of the three conjectures mentioned above:
 the asymptotic (\ref{defccb}), the classical bootstrap equations
 (\ref{clasboot}), and the relation (\ref{glengthII}).
Up to our knowledge no
rigorous proof of any of these relations is known.
The basic difficulty is the conformal block itself which except of some special cases
is only known as a formal power series in the $x$ variable.
Since the coefficients are defined in terms
of inverses of the Gram matrices of
a Verma module
a direct analysis is prohibitively difficult.

A more efficient method based on a recurrence relation for the coefficients
was developed by Al.~B.~Zamolodchikov \cite{Zamolodchikov:ie}.
The assumption crucial for this method
is that the formal power series defining the conformal block converges.
It is actually believed (and well justified by all special cases
where the conformal block can be calculated explicitly)
that the radius of convergence is 1.
Another commonly accepted hypothesis, still waiting for its rigorous  proof,
states that the only singularities of the conformal block with
respect to the $z$ variable are branching points (in general of a transcendental kind)
at $0$, $1$, and $\infty$ \cite{ZZ}. This implies
that the conformal block is a single-valued analytic function on the
universal covering of a 3-punctured Riemann sphere and can
be expressed  by a power series convergent in the entire domain
of its analyticity \cite{Zam}. A recurrence relation for
calculating coefficients of this so called $q$-expansion
\cite{Zam} provides an extremely efficient method
for numerical analysis of conformal block and was applied for
testing the conformal bootstrap equations
\cite{Zamolodchikov:1995aa,Runkel:2001ng}. Let us note that both the $x$- and $q$-expansion
recurrence methods involve formulae which successfully passed numerical tests but
a formal proof of them is still lacking.

The organization of the paper is as follows.
In Sect.\ 2 we briefly present the classical Liouville action and its relation to the monodromy
problem of Fuchsian uniformization.
In Sect.\ 3 the classical conformal block is introduced and its relation to a certain
monodromy problem of the null-vector decoupling equation is clarified.
These sections contain a material which is basically known
and were added mainly for completeness
and to set the notation.

In Sect.\ 4 the Zomolodchikov's recurrence method is applied
to calculate the $x$- and $q$-expansion of the classical conformal block in the case
of parabolic external weights up to 7th and 16th power, respectively.
It is also checked by symbolic calculation that up to these orders the asymptotic
(\ref{defccb}) is correct.

In Sect.\ 5 the saddle point weights and momenta are defined and numerically calculated
in same special cases in which the lengths of geodesics in each channel are exactly known.
An excellent agreement is obtained using both
the $x$- and $q$-expansions of classical conformal block.
It is also verified that the geodesic length function numerically calculated from the
formula (\ref{glengthII}) satisfies the upper and lower bounds known to mathematicians \cite{Hempel}.
The calculations of this section provide a strong evidence that the formula
(\ref{glengthII}) is correct and along with the $q$-expansion for classical conformal block
give an extremely efficient method for numerical calculations of the geodesic length
functions on the 4-punctured sphere.

In Sect.\ 6 some  numerical tests of the classical bootstrap are presented.
It is shown that it is satisfied with high precision, improving with the number of
terms taken into account in the $q$-expansion of the conformal block.
This allows for calculating the classical Liouville action for 4-puncture sphere
and, by the Polyakov conjecture, the accessory parameters of the Fuchsian uniformization.
Finally, Sect.~7 contains conclusions and discussion of
possible extensions of the present work.

\section{Classical Liouville action and the Polyakov conjecture}
\setcounter{equation}{0}

A conformal factor of the hyperbolic metric on the surface $X$ parametrized
with a complex coordinate $z$ is a solution of the  Liouville equation
\begin{equation}
\label{Liouville}
\partial_z\partial_{\bar z} \phi(z,\bar z) =
\frac{\mu}{2} {\rm e}^{\phi(z,\bar z)}
\end{equation}
Given a genus $g$ of $X$ and a set of points
$z_1,\ldots, z_n$ removed from $X$
this metric is determined by the singular behavior of $\phi$ at $z_j-$s.
In the present paper we shall consider the case of $X$ being a punctured sphere
($g = 0$) and choose a complex coordinates on $X$ in such a way that $z_n = \infty.$
The  existence and the uniqueness of the solution of the
equation  (\ref{Liouville}) on the sphere with the elliptic singularities,
\begin{equation}
\label{asymptot:elliptic}
\phi(z,\bar z) = \left\{
\begin{array}{lll}
-2\left(1-\xi_j \right)\log | z- z_j |  + O(1) & {\rm as } & z\to z_j,
\hskip 5mm j = 1,\ldots,n-1,\\
-2\left(1+\xi_n \right)\log | z| + O(1) & {\rm as } & z\to \infty,
\end{array}
\right.
\end{equation}
was proved by Picard \cite{Picard1,Picard2} (see also \cite{Troyanov} for a modern proof).
The solution can be interpreted as a conformal factor of the complete, hyperbolic metric on
\(
X = {\mathbb C}\setminus\{z_1,\ldots,z_{n-1}\}
\)
with the conical singularities
of the opening angles $0<2\pi\xi_j<2\pi$
at the punctures $z_j$. The solution is known to exist also in
the case of parabolic singularities (corresponding to $\xi_j \to 0$),
with the asymptotic behavior of the Liouville field of
the form
\begin{equation}
\label{asymptot:parabolic}
\phi(z,\bar z) = \left\{
\begin{array}{lll}
-2\log |z- z_j |  -2\log \left|\log |z- z_j |\right| + O(1) & {\rm as } & z\to z_j, \\
-2\log |z| - 2\log \left|\log |z|\right| + O(1) & {\rm as } & z\to \infty.
\end{array}
\right.
\end{equation}

The central objects in the geometric approach to the quantum Liouville
theory are the partition functions on $X\!:$
\begin{equation}
\label{partition}
\left\langle X \right\rangle =
\int\limits_{\cal M}{\cal D}\phi\;{\rm e}^{-Q^2 S_{\rm L}[\phi]},
\end{equation}
where $\cal M$ is the space of conformal factors on $X$ with the asymptotics
(\ref{asymptot:elliptic}) or (\ref{asymptot:parabolic}),
and the correlation functions of the energy--momentum tensor
\begin{equation}
\label{emom}
\left\langle
T(u_1)\ldots T(u_k) \bar T(\bar w_1)\ldots\bar T(\bar w_l)\, X
\right\rangle  =
\int\limits_{\cal M}\!\!{\cal D}\phi\;{\rm e}^{-Q^2 S_{\rm L}[\phi]}\
T(u_1)\ldots T(u_k) \bar T(\bar w_1)\ldots\bar T(\bar w_l),
\end{equation}
with
\begin{equation}
\label{emom:def}
T(u) =  \frac{Q^2}{2}\left[
-\frac12\left(\partial_u\phi(u,\bar u)\right)^2 + \partial^2_u\phi(u,\bar u)
\right].
\end{equation}

The singular nature of the Liouville field at the punctures
requires regularizing terms in the Liouville action:
\begin{eqnarray}
\label{action}
\nonumber
S_{\rm L}[\phi]
& = &
\frac{1}{4\pi}
\lim_{\epsilon\to 0}
S_{\rm L}^\epsilon[\phi], \\
S_{\rm L}^\epsilon[\phi]
& = &
\int\limits_{X_\epsilon}\!d^2z
\left[\left|\partial\phi\right|^2 + \mu{\rm e}^{\phi}\right]
+  \sum\limits_{j=1}^{n-1}\left(1-\xi_j\right)
\hspace{-4mm}\int\limits_{|z-z_j|=\epsilon}\hspace{-4mm}|dz|\ \kappa_z \phi
+\left(1+\xi_n\right)
\hspace{-2mm}\int\limits_{|z|=\frac{1}{\epsilon}}\hspace{-2mm}|dz|\ \kappa_z \phi
\\
\nonumber
&&
- 2\pi\sum\limits_{j=1}^{n-1}\left(1-\xi_j\right)^2\log\epsilon
- 2\pi\left(1+\xi_n\right)^2\log\epsilon,
\end{eqnarray}
where \( X_\epsilon = {\mathbb C}\setminus\left\{\left(\bigcup_{j=1}^n |z-z_j|< \epsilon\right)
\cup \left(|z|>\frac{1}{\epsilon}\right)\right\}. \) The form of
(\ref{action}) is valid for parabolic singularities (with
corresponding $\xi_j = 0$) as well.

One can check by perturbative calculations of the correlators (\ref{emom})
\cite{Takhtajan:yk} that
 the central charge reads
\begin{equation}
\label{centrsl:charge}
c = 1 + 6Q^2.
\end{equation}
The transformation properties of (\ref{partition}) with respect to the global conformal
transformations show \cite{Takhtajan:yk} that the punctures behave like
primary fields with the dimensions
\begin{equation}
\label{Delta}
\Delta_j =
\bar{\Delta}_j = \frac{Q^2}{4}\left(1-\xi_j^2\right).
\end{equation}
As for fixed $\xi_j$ the dimensions scale like $Q^2$,
the punctures correspond to {\em heavy} fields
of the operator approach \cite{Zamolodchikov:1995aa}.

In the classical limit $Q^2 \to \infty$ with all
\begin{equation}
\label{small:deltas}
\delta_i  \stackrel{\rm def}{=}   \frac{\Delta_i}{Q^2} = \frac{1-\xi_j^2}{4}
\end{equation}
kept fixed  we expect the path integral to be
dominated by the classical action:
\begin{equation}
\label{asymptotic:X}
\left\langle X \right\rangle
\sim
{\rm e}^{-Q^2S^{\rm (cl)}(\delta_i\,;\,z_i)},
\end{equation}
where the classical action $S^{\rm (cl)}(\delta_i\,;\,z_i)$ is the
functional $S_{\rm L}[\,\cdot\,]$ (\ref{action}) evaluated at the
classical solution
 $\varphi$  of (\ref{Liouville}) with the asymptotics
(\ref{asymptot:elliptic}) or (\ref{asymptot:parabolic}).
Similarly:
\begin{equation}
\label{classical:withT}
\left\langle T(z) X \right\rangle
\sim
T^{\rm cl}(z)\ {\rm e}^{-Q^2 S^{\rm (cl)}(\delta_i\,;\,z_i)},
\end{equation}
where $T^{\rm cl}(z)$ is the classical energy--momentum tensor.

From (\ref{emom:def}) and (\ref{asymptot:elliptic}) or
(\ref{asymptot:parabolic}) it follows that
\begin{eqnarray}
\label{singular:behavior}
\nonumber
T^{\rm cl}(z) & \sim &  \frac{\Delta_j}{(z-z_j)^2}
 \hskip 1cm  {\rm for}\;\; z \to z_j, \\
T^{\rm cl}(z) & \sim &   \hskip .5cm \frac{\Delta_n}{z^2}
 \hskip 1.5cm  {\rm for}\;\; z \to \infty,
\end{eqnarray}
and consequently
\begin{equation}
\label{T:classical}
T^{\rm cl}(z)  =
Q^2 \sum\limits_{j=1}^{n-1}
\left[\frac{\delta_j}{(z-z_j)^2} + \frac{c_j}{z-z_j}\right].
\end{equation}
Combining  now (\ref{asymptotic:X}), (\ref{classical:withT})
and (\ref{T:classical})
with the conformal Ward identity \cite{Belavin:1984vu}
\begin{equation}
\label{Ward}
\left\langle T(z) X\right\rangle =
\sum\limits_{j=1}^{n-1}
\left[\frac{\Delta_j}{(z-z_j)^2} + \frac{1}{z-z_j}\frac{\partial}{\partial z_j}\right]
\left\langle X \right\rangle,
\end{equation}
we get the relation
\begin{equation}
\label{PC}
c_j  =  - \frac{\partial S^{\rm (cl)}(\delta_i\,;\,z_i)}{\partial z_j}
\end{equation}
known as the  Polyakov conjecture.

It is amazing that this relation obtained by general heuristic
path integral arguments turned out to provide an exact solution to a long
standing problem of the so called accessory parameters of the Fuchsian
uniformization of the punctured sphere.
Indeed, it was rigorously proved
\cite{ZoTa1,ZoTa2,Zograf,Cantini:2001wr,Takhtajan:2001uj,Hadasz:2003kp}
that the formula
(\ref{PC}) yields the accessory parameters $c_j$
for which
the Fuchsian equation
\begin{equation}
\label{Fuchs}
\partial^2\psi(z) + \frac{1}{Q^2}T^{\rm cl}(z)\psi(z) = 0
\end{equation}
admits a fundamental system of solutions
with  $SU(1,1)$ monodromies around all singularities. Note that
if $\{\chi_1(z), \chi_2(z)\}$ is such a system
then the function $\varphi(z,\bar z)$ determined by the relation
\begin{equation}
\label{phi}
e^{\varphi(z,\bar z)}  =
{4\,|w'|^2 \over \mu (1 - |w|^2 )^2},
\hskip 5mm
w(z) = {\chi_1(z)\over\chi_2(z)},
\end{equation}
satisfies (\ref{Liouville}) and (\ref{asymptot:elliptic}) (or (\ref{asymptot:parabolic})).
The $SU(1,1)$ monodromy condition is then equivalent to the existence of the
well defined hyperbolic metric on $X.$

From (\ref{singular:behavior}) it follows that $c_j$ satisfy the
relations
\begin{equation}
\label{restrictions}
\sum\limits_{j=1}^{n-1}c_j  =  0,
\hskip 1cm
\sum\limits_{j=1}^{n-1}\left(\delta_j + z_jc_j\right)  =  \delta_n.
\end{equation}
In the case of three singularities the only two accessory parameters
are completely determined by (\ref{restrictions}). Hence
one can solve the equation (\ref{Fuchs}), find the classical
Liouville field and calculate the classical action. The result
is \cite{Zamolodchikov:1995aa}:
\begin{eqnarray}
\label{action:elliptic:3}
S^{\rm (cl)}(\delta_3,\delta_2,\delta_1)& \equiv & S^{\rm (cl)}(\delta_3,\delta_2,\delta_1;\infty,1,0) =
\\
\nonumber
&&
\hspace*{-2cm}
\frac12(1-\xi_1-\xi_2-\xi_3)\log\mu +
\!\!\!\sum\limits_{\sigma_2,\sigma_3 = \pm}\!\!\!\!
F\left(\frac{1-\xi_1}{2} + \sigma_2\frac{\xi_2}{2} +\sigma_3\frac{\xi_3}{2}\right)
-\sum\limits_{j=1}^3F(\xi_j)
+ {\rm const},
\end{eqnarray}
where
\[
F(x)  =  \int\limits_{1/2}^{x}\!dy\; \log\frac{\Gamma(y)}{\Gamma(1-y)}.
\]
For the elliptic and parabolic singularities one has $\delta \leqslant \frac14.$
As we shall see below, it is useful to know the classical Liouville action also
for $\delta > \frac14,$ what corresponds to the hyperbolic singularities.
The relevant construction of the metric and the classical action was given in \cite{Hadasz:2003he}.
If we write
\begin{equation}
\label{paramet}
\delta_3  =  \frac14 + p^2,
\hskip 1cm
p \in {\mathbb R},
\end{equation}
then in the case of two elliptic/parabolic and one hyperbolic
singularity
\begin{eqnarray}
\label{action:mixed:3}
\nonumber
S^{\rm ( cl)}(\delta_3,\delta_2,\delta_1)
& = &
\frac12(1-\xi_1-\xi_2)\log\mu +
\!\!\!\!\!\sum\limits_{\sigma_2,\sigma_3 = \pm}\!\!\!\!
F\left(\frac{1-\xi_1}{2} + \sigma_2\frac{\xi_2}{2} +i \sigma_3p\right)
\\
&&-\sum\limits_{j=1}^2F(\xi_j)
 + H(2i p) + \pi|p| +{\rm const},
\end{eqnarray}
with
\[
H(x)  =  \int\limits_0^x\!dy\;\log\frac{\Gamma(-y)}{\Gamma(y)}.
\]

\section{Classical conformal block}
\setcounter{equation}{0}

The partition function (\ref{partition}) corresponds in the
operator formulation to the correlation function of the primary
fields $V_{\alpha_j}(z_j,\bar z_j),$
\begin{equation}
\label{relation} \left\langle X \right\rangle  =
\Big\langle
V_{\alpha_n}(\infty,\infty)\ldots V_{\alpha_1}(z_1,\bar z_1)
\Big\rangle,
\end{equation}
where
\[
\Delta_j = \alpha_j(Q-\alpha_j),\;\;\;
\alpha_j = \frac{Q}{2}\left(1 + \xi_j\right).
\]
The DOZZ 4-point correlation function with the standard locations
$z=0,x,1,\infty$
is expressed as an integral over the continuous spectrum
\begin{eqnarray}
\label{four:point:}
&&
\hspace*{-2.5cm}
\Big\langle
V_{\alpha_4}(\infty,\infty)V_{\alpha_3}(1,1)V_{\alpha_2}(x,\bar x)V_{\alpha_1}(0, 0)
\Big\rangle =
\\
\nonumber
&& \int\limits_{\frac{Q}{2} + i{\mathbb R}^{+}}\!\!\!\!\!\!\!d\alpha\;
C(\alpha_4,\alpha_3,\alpha)C(Q-\alpha,\alpha_2,\alpha_1) \left|
{\cal F}_{1+6Q^2,\Delta}\left[^{\Delta_3\ \Delta_2}_{\Delta_4\
\Delta_1}\right](x) \right|^2.
\end{eqnarray}
Let
\[
\mbox{\bf 1}_{\Delta,\Delta} = \sum_{I}
(\ket{\xi_{\Delta,I}}\otimes \ket{\xi_{\Delta,I}})
(\bra{\xi_{\Delta,I}}\otimes\bra{\xi_{ \Delta,I}})
\]
be an operator that projects onto the space spanned by the states
form the conformal family with the highest weight $\Delta$. The
correlation function with the $\mbox{\bf 1}_{\Delta,\Delta} $
insertion factorizes into the product of the holomorphic and
anti-holomorphic factors,
\begin{eqnarray}
\label{c4}
&&
\hspace*{-1cm}
 \Big\langle
V_4(\infty,\infty)V_3(1,1)\mbox{\bf 1}_{\Delta,\Delta}
V_2(x,\bar x)V_1(0,0)
\Big\rangle  =  \\
 \nonumber
&& C(\alpha_4,\alpha_3,\alpha)\,C(Q-\alpha,\alpha_2,\alpha_1)\,
{\cal F}_{\!1+6Q^2,\Delta}
\!\left[_{\Delta_{4}\;\Delta_{1}}^{\Delta_{3}\;\Delta_{2}}\right]
\!(x)\,
{\cal F}_{\!1+6Q^2,\Delta}
\!\left[_{\Delta_{4}\;\Delta_{1}}^{\Delta_{3}\;\Delta_{2}}\right]
\!(\bar x).
\end{eqnarray}
Assuming a path integral representation of the l.h.s. one should expect
in the limit $Q \to \infty$, with all the weights being heavy $\Delta,\Delta_i \sim Q^2$,
the  following asymptotic behavior
\begin{equation}
\label{a4}
\Big\langle
V_4(\infty,\infty)V_3(1,1)\mbox{\bf 1}_{\Delta,\Delta}
V_2(x,\bar x)V_1(0,0)
 \Big\rangle \sim
{\rm e}^{-Q^2
S^{\rm (cl)}(\delta_i,x;\delta)
}.
\end{equation}
On the other hand one can calculate this limit for the DOZZ coupling constants
\cite{Zamolodchikov:1995aa,Hadasz:2003he} obtaining
\begin{equation}
\label{asymptotC}
C(\alpha_4,\alpha_3,\alpha)C(Q-\alpha,\alpha_2,\alpha_1)
 \sim
{\rm e}^{-Q^2\left(
S^{\rm cl}(\delta_4,\delta_3,\delta)
+ S^{\rm cl}(\delta,\delta_2,\delta_1)
\right)}.
\end{equation}
It follows that the conformal block
should have the $Q\to \infty$ asymptotic (\ref{defccb})
so that
\begin{equation}
\label{deltaaction}
S^{\rm (cl)}(\delta_i,x;\delta)=S^{\rm (cl)}(\delta_4,\delta_3,\delta)
+ S^{\rm (cl)}(\delta,\delta_2,\delta_1)-
f_{\delta}
\!\left[_{\delta_{4}\;\delta_{1}}^{\delta_{3}\;\delta_{2}}\right](x)
-\bar f_{\delta}
\!\left[_{\delta_{4}\;\delta_{1}}^{\delta_{3}\;\delta_{2}}\right](\bar x).
\end{equation}
It should be stressed that the asymptotic behavior (\ref{defccb}) is a nontrivial
statement on the (quantum) conformal block. Although there
is no proof of this property yet it seems to be well justified by
sample numerical calculations, as well as by its consequences.
We shall briefly describe two of them.

The first one is the classical bootstrap mentioned in the introduction.
In the semiclassical limit the l.h.s of the formula (\ref{four:point:})
takes the form ${\rm e}^{-Q^2 S^{\rm (cl)}(\delta_4,\delta_3,\delta_2,\delta_1;x)}$,
where
$$
S^{\rm (cl)}(\delta_4,\delta_3,\delta_2,\delta_1;x)
\equiv
S^{\rm (cl)}(\delta_4,\delta_3,\delta_2,\delta_1;\infty,1,x,0)\ .
$$
The r.h.s. of (\ref{four:point:}) is in this limit determined by the saddle point approximation
$$
 {\rm e}^{-Q^2S^{\rm (cl)}(\delta_i,z_i;\delta_s)}
\; \approx \;
 \int\limits_0^\infty\!dp\; {\rm e}^{-Q^2S^{\rm (cl)}(\delta_i,x;\delta)}
$$
where
$
\delta_s ={\textstyle {1\over 4}} +p_s^2
$
and the saddle point Liouville momentum $p_s$ is determined by
\begin{equation}
\label{saddle} {\partial \over \partial p}S^{\rm
(cl)}(\delta_i,x;{\textstyle {1\over 4}} +p_s^2)_{|p=p_s}=0.
\end{equation}
One thus gets the relation (\ref{clasfact}) first obtained in
\cite{Zamolodchikov:1995aa} and the classical bootstrap
(\ref{clasboot}) as its consistency condition.

The second implication of the asymptotic (\ref{defccb}) is the relation of the classical block
to  certain  monodromy problem, which in fact proved
to be essential for developing a very effective recursive method for calculating the
conformal block itself (and therefore also its classical asymptotic) \cite{Zam,Zamolodchikov:ie,ZZ}.
Consider the null vectors on the second level of the Verma module, given by
\begin{equation}
\label{rel1}
\ket{\chi_{\pm}}  =  \left(L_{-2} - \frac{3}{2(2\Delta_{\pm} +1)}\,L_{-1}^2\right)\ket{\Delta_\pm},
\end{equation}
where $\ket{\Delta_\pm}$ are the highest weight states with (for $c \geq 25$)
\[
\Delta_{\pm}  =
\frac{1}{16}\left(5-c\pm\sqrt{(c-1)(c-25)}\right).
\]
This expression simplifies if we use convenient parametrization of the central charge
\begin{equation}
\label{convenient}
c=1+6Q^2, \;\;\;\;\;\;Q={\textstyle{1\over b}}+b,
\end{equation}
so that
\[
\Delta_{+} = -\frac12 -\frac34 b^2 ,
\hskip 1cm
\Delta_{-} = -\frac12 -\frac3{4 b^2}.
\]
It follows
from (\ref{rel1})  that the correlators
\begin{equation}
\label{chi}
\left\langle\hat\chi_{\pm}(z) X\right\rangle_\Delta
\; \stackrel{\rm def}{=}  \;
\Big\langle
V_4(\infty,\infty)V_3(1,1)\mbox{\bf 1}_{\Delta,\Delta}
\hat\chi_{\pm}(z) V_2(x,\bar x)V_1(0,0) \Big\rangle,
\end{equation}
where $\hat\chi_\pm(z)$ are the null fields corresponding to $\ket{\chi_{\pm}},$
satisfy the null-vector decoupling equation:
\begin{eqnarray}
\label{Fuchs1}
&&
\left[
\frac{\partial^2}{\partial z^2}
+\gamma_\pm\left(\frac{1}{z} - \frac{1}{1-z}\right)\frac{\partial}{\partial z}
\right]
\left\langle\hat\chi_{\pm}(z) X\right\rangle_\Delta =
 \\
\nonumber
&&
\gamma_\pm
\left[\frac{\Delta_1}{z^2} + \frac{\Delta_2}{(z-x)^2}
+  \frac{\Delta_3}{(1-z)^2} +
\frac{\Lambda_{\pm}}{z(1-z)}
+ \frac{x(1-x)}{z(z-x)(1-z)}
\frac{\partial}{\partial x}\right]
 \left\langle\hat\chi_{\pm}(z) X\right\rangle_\Delta,
\end{eqnarray}
with
\[
\Lambda_\pm  =
\Delta_1 + \Delta_2 + \Delta_3 + \Delta_\pm - \Delta_4
\]
and
\[
\gamma_\pm = \frac23(2\Delta_\pm + 1)
\hskip 1cm
\Rightarrow
\hskip 1cm
\gamma_+ = -b^2, \hskip 5mm \gamma_- = -\frac{1}{b^2}.
\]
For $Q\to \infty$ we have either $b\to 0$ or $b \to \infty.$ To
fix the notation we shall concentrate on the first possibility.
For $b \to 0$ the operator with the weight $\Delta_+$ remains
``light'' ($\Delta_+ = {\cal O}(1)$) and its presence in the
correlation function has no influence on the classical solution of
the field equations\footnote{
This is reflected by the fact that limit $b\to 0$ of the equation
(\ref{Fuchs1}) exists  only in the ``+'' case.
}.
Consequently, for $b\to 0$
\begin{equation}
\label{a5}
\left\langle\hat\chi_{+}(z) X\right\rangle_\Delta \sim
\chi^{\rm cl}(z)\,
{\rm e}^{-{1\over b^2}\left(
S^{\rm (cl)}(\delta_4,\delta_3,\delta) +
S^{\rm (cl)}(\delta,\delta_2,\delta_1)
-f_{\delta}\!\left[_{\delta_{4}\;\delta_{1}}^{\delta_{3}\;\delta_{2}}\right](x)
-\bar f_{\delta}\!\left[_{\delta_{4}\;\delta_{1}}^{\delta_{3}\;\delta_{2}}\right](\bar x)\right)}
\end{equation}
and we get from (\ref{Fuchs1}):
\begin{equation}
\label{Fuchs2}
\partial^2_z\chi^{\rm cl}(z) +
\left[\frac{\delta_1}{z^2} + \frac{\delta_2}{(z-x)^2}
+  \frac{\delta_3}{(1-z)^2}
+ \frac{\delta_1+\delta_2 + \delta_3 - \delta_4}{z(1-z)}
+ \frac{x(1-x){\cal C}(x)}{z(z-x)(1-z)}\right]\chi^{\rm cl}(z)  = 0,
\end{equation}
where
\begin{equation}
\label{accessory2}
{\cal C}(x) =  {d\over d x}\,
f_{\delta}\!\left[_{\delta_{4}\;\delta_{1}}^{\delta_{3}\;\delta_{2}}\right]\!(x).
\end{equation}

The accessory parameter ${\cal C}(x)$ can be determined from the following monodromy problem.
Consider the correlation function
\begin{equation}
\label{f:def}
G(z) \; \stackrel{\rm def}{=} \;
\Big\langle V_4(\infty,\infty)V_3(1,1)\hat\chi_{+}(z)V_\Delta(0,0)\Big\rangle
\end{equation}
where $V_\Delta$ is the primary field corresponding to the highest weight
state $\xi_\Delta.$ The null vector decoupling equation for this correlator reads
\begin{equation}
\label{Fuchs:s}
\left(
\frac{1}{b^2}\frac{d^2}{dz^2}
+ \left(\frac{1}{1-z} - \frac{1}{z}\right)\frac{d}{dz}
+\frac{\Delta\,}{z^2} + \frac{\Delta_3}{(1-z)^2}
+ \frac{\Delta + \Delta_+ + \Delta_3 - \Delta_4}{z(1-z)}
\right) G(z) = 0.
\end{equation}
Substituting into this equation the most singular term
in the OPE
\[
\hat\chi_{+}(z)V_\Delta(0,\bar 0)  \sim  z^\kappa\, V_{\Delta'}(0,\bar 0)
\]
we get
\begin{equation}
\label{kappa}
\kappa(\kappa-1) +b^2(\Delta - \kappa) = 0.
\end{equation}
In the  limit $b\to 0:$
\[
\Delta =
\frac{Q^2}{4}(1-\xi^2) =
\frac{1}{4b^2}(1-\xi^2) + {\cal O}(1)
\]
and (\ref{kappa}) transforms to
\[
\kappa(\kappa -1) +\frac14(1-\xi^2) = 0
\hskip 1cm
\Rightarrow
\hskip 1cm
\kappa = \frac12(1\pm\xi).
\]
For $z \;\to\;  {\rm e}^{2\pi i}\,z:$
\[
\left(
\begin{array}{c}
z^{\frac{1-\xi}{2}} \\
z^{\frac{1+\xi}{2}}
\end{array}
\right)
\; \to \;
-\left(
\begin{array}{cc}
{\rm e}^{-i\pi\xi} & 0 \\
0 & {\rm e}^{i\pi\xi}
\end{array}
\right)
\left(
\begin{array}{c}
z^{\frac{1-\xi}{2}} \\
z^{\frac{1+\xi}{2}}
\end{array}
\right)
\]
and (minus) the trace of the monodromy matrix (an invariant with
respect to the choice of the basis in the space of solutions of
(\ref{Fuchs:s})) is equal to
\[
2\cos\pi\xi.
\]

Note that we get the same monodromy invariant if we replace in
(\ref{f:def}) the primary field $V_{\Delta}$ with any of its
descendants. Note also that (in the classical limit) the monodromy invariant
of the two independent solutions of (\ref{Fuchs:s})
for a curve encircling $0$ is by construction equal to the monodromy
invariant for a basis in the space of solutions of (\ref{Fuchs2}) along
a curve encircling both $0$ and $x$ (all point on this curve can be
taken ``to the left'' of the operator $\mbox{\bf 1}_\Delta$). This
condition fixes the accessory parameter ${\cal C}(x)$ that appears on
the r.h.s. of (\ref{Fuchs2}).

\section{Zamolodchikov's recursion  methods}
\setcounter{equation}{0}

The BPZ 4-point conformal block is defined \cite{Belavin:1984vu}
as a formal power series\footnote{It is believed that the series (\ref{QQblock}) converges on the unit disc but
up to our knowledge there is no rigorous proof of this fact yet.}
\begin{equation}
\label{QQblock}
{\cal
F}_{\!c,\Delta}\!\left[_{\Delta_{4}\;\Delta_{1}}^{\Delta_{3}\;\Delta_{2}}\right]\!(\,x)
= x^{\Delta-\Delta_{2}-\Delta_{1}}\left( 1 + \sum_{n=1}^\infty
x^{\;n} {\cal
F}^{\,n}_{\!c,\Delta}\!\left[_{\Delta_{4}\;\Delta_{1}}^{\Delta_{3}\;\Delta_{2}}\right]
\right).
\end{equation}
Studying the analytic structure of its coefficients
Al.\ Zamolodchikov derived the
 recursion relation \cite{Zamolodchikov:ie}:
\begin{eqnarray}
\label{recrel}
{\cal
F}^{\,n}_{\!c,\Delta}\!\left[_{\Delta_{4}\;\Delta_{1}}^{\Delta_{3}\;\Delta_{2}}\right]
&=&{g}^{\,n}_{\Delta}\!\left[_{\Delta_{4}\;\Delta_{1}}^{\Delta_{3}\;\Delta_{2}}\right]
+\!\!\!\!
\sum_{\begin{array}{c}\scriptstyle r\geq 2\; s\geq 1\\[-3pt]\scriptstyle n\,\geq \,rs \,\geq 2 \end{array}}
\!\!\!\!\! \frac{\tilde
R^{\,rs}_\Delta\!\left[_{\Delta_{4}\;\Delta_{1}}^{\Delta_{3}\;\Delta_{2}}\right]}{c-c_{rs}(\Delta)}
\;{\cal
F}^{\,n-rs}_{\!c_{rs}(\Delta),\Delta+rs}\!\left[_{\Delta_{4}\;\Delta_{1}}^{\Delta_{3}\;\Delta_{2}}\right],
\end{eqnarray}
where
\begin{eqnarray*}
\nonumber
c_{rs}(\Delta)&=&  13 - 6 \left(T_{rs}(\Delta)+{1\over T_{rs}(\Delta)}\right),\\
T_{rs}(\Delta)&=&\textstyle \frac{rs -1 + 2\,\Delta  +
    {\sqrt{{\left( r - s \right) }^2 +
        4\,\left(r\,s -1 \right) \,\Delta  +
        4\,{\Delta }^2}}}{r^2-1},
\nonumber
\end{eqnarray*}
and
${g}^{\,n}_{\Delta}\!\left[_{\Delta_{4}\;\Delta_{1}}^{\Delta_{3}\;\Delta_{2}}\right]$
are coefficients of the expansion of the hypergeometric function,
$$
{}_2F_1(\Delta +\Delta_2-\Delta_1,\Delta +\Delta_3 -\Delta_4,
2\Delta,x) =\sum\limits_{n=0}^\infty
{g}^{\,n}_{\Delta}\!\left[_{\Delta_{4}\;\Delta_{1}}^{\Delta_{3}\;\Delta_{2}}\right]x^n.
$$
An exact form of the coefficients $\tilde
R^{\,rs}_\Delta\!\left[_{\Delta_{4}\;\Delta_{1}}^{\Delta_{3}\;\Delta_{2}}\right]$ (see Appendix)
was partially
derived and partially guessed in \cite{Zamolodchikov:ie}. Although no proof of this form exists
it is well justified by numerical calculations.

Once the expansion (\ref{QQblock}) is known  one can calculate
the coefficients of the power expansion of the classical conformal block
\begin{equation}
\label{classblock}
f_{\delta}\!\left[_{\delta_{4}\;\delta_{1}}^{\delta_{3}\;\delta_{2}}\right]\!(\,x)
= (\delta-\delta_1-\delta_2) \log x +  \sum_{n=1}^\infty
x^{\;n} f^{\,n}_{\delta}\!\left[_{\delta_{4}\;\delta_{1}}^{\delta_{3}\;\delta_{2}}\right]
\end{equation}
directly from the asymptotic (\ref{defccb})
\begin{equation}
\label{limit} \sum_{n=1}^\infty x^{\;n}
f^{\,n}_{\delta}\!\left[_{\delta_{4}\;\delta_{1}}^{\delta_{3}\;\delta_{2}}\right]
= \lim\limits_{Q^2 \to \infty} {1\over Q^2} \log\left(1 +
\sum_{n=1}^\infty x^{\;n} {\cal
F}^{\,n}_{\!c,\Delta}\!\left[_{\Delta_{4}\;\Delta_{1}}^{\Delta_{3}\;\Delta_{2}}\right]
\right),
\end{equation}
where on the r.h.s.\ one first expand the logarithm into a
 power series and then the limit is taken for each term separately.

In the present work we were interested in a special case
of  all parabolic external weights $\Delta_i={Q^2\over 4}$, the central charge $c=1+6Q^2$,
and the intermediate weight $\Delta=Q^2\left({1\over 4}+p^2\right)$ parameterized by the Liouville momenta $p$.
We have checked  by symbolic  calculations that up to $m=7$
the limits in (\ref{limit}) exist yielding explicit formulae for the first seven coefficients
of the expansion (\ref{classblock}). Up to the first five terms:
\begin{eqnarray}
\label{x-expansion}
f(p,x)&\equiv &\textstyle
f_{{1\over 4} +p^2 }\!\left[\;_{{1\over 4}\;\;\;{1\over 4}}^{{1\over 4}\;\;\;{1\over 4}}\;\right]\!(\,x)
\\ \nonumber
&=&\textstyle
\left(p^2 -{1\over 4}\right)\log x +\left({1\over 8}+{p^2\over 2}\right)x
\\ \nonumber
&+&\textstyle
\left( {9\over 128} + {13\ p^2\over 64} + {1\over 1024\ (1 + p^2)}\right)x^2
\\ \nonumber
&+&\textstyle
\left( {19\over 384} + {23\ p^2\over 192} + {1\over 1024\ (1 + p^2)}\right)x^3
\\ \nonumber
&+&\textstyle
\left( {1257\over 32768} + {2701\ p^2\over 32768}- {1\over 2097152\ (1 + p^2)^3}\right.
\\ \nonumber
    &&\textstyle
    \left. +
    {3\over 8388608\ \left(1 + p^2\right)^2}+ {7439\over 8388608\ (1 + p^2)} +
    {81\over 8388608\ (4 + p^2)}\right)x^4
    \\ \nonumber
& + &\textstyle
\left({2573\over 81920} + {5057\ p^2\over 81920}- {1\over 1048576\ (1 + p^2)^3}\right.
\\ \nonumber
    &&\textstyle
    \left. +
    {3\over 4194304 \ \left(1 + p^2\right)^2} + {3343\over 4194304\ (1 + p^2)} +
    {81\over 4194304\ (4 + p^2)}\right) x^5 +\dots
\end{eqnarray}
The limitation of the formulae (\ref{QQblock}) and (\ref{classblock})  is that the power series
involved are supposed to converge only for $|x|<1$.

A more convenient representation of the conformal block
was developed  by Al.\ Za\-mo\-lod\-chi\-kov \cite{Zam} who proposed to regard it as a function of
the variable
$$
q(x) = {\rm e}^{-\pi {K(1-x)\over K(x)}},
\hskip 1cm
K(x)=\int\limits_{0}^1{dt\over \sqrt{(1-t^2)(1-xt^2)}}.
$$
The map $\mathbb{C}\setminus \{0,1,\infty\} \ni x\to q(x)\in \mathbb{D}$
yields a uniformization of the 3-punctured sphere by the Poincar\'e disc $\mathbb{D}$. If the points
$0,1,\infty$ are the only singular points of the conformal block, then the block is a single
valued function on $\mathbb{D}$ and the series in $q$ variable converges uniformly
on each subset $\{q:|q|<{\rm e}^{-\epsilon}<1\}$. It was shown in \cite{Zam} that the conformal
block can be expressed as
\begin{eqnarray}
\nonumber
{\cal F}_{\!c,\Delta}\!\left[_{\Delta_{4}\;\Delta_{1}}^{\Delta_{3}\;\Delta_{2}}\right]\!(\,x)
&=&
x ^{{c-1\over 24}-\Delta_1-\Delta_2}
(1-x)^{{c-1\over 24}-\Delta_1-\Delta_3}\\
\nonumber
&\times &
\theta_3(q)^{{c-1\over 2}-4(\Delta_1+\Delta_2+\Delta_3+\Delta_4)}
\\[4pt]
\label{Hrep}
&\times &
(16 q)^{\Delta - {c-1\over 24}} H_{\!c,\Delta}\!\left[_{\Delta_{4}\;\Delta_{1}}^{\Delta_{3}\;\Delta_{2}}\right]\!(\,q)\\[6pt]
\nonumber
&=&
x ^{{c-1\over 24}-\Delta_1-\Delta_2}
(1-x)^{{c-1\over 24}-\Delta_1-\Delta_3}\\
\nonumber
&\times &
\left( {\textstyle{2\over\pi}}K(x)\right)^{{c-1\over 4}-2(\Delta_1+\Delta_2+\Delta_3+\Delta_4)}\\[4pt]
\nonumber
&\times&
(16 q)^{\Delta - {c-1\over 24}}H_{\!c,\Delta}\!\left[_{\Delta_{4}\;\Delta_{1}}^{\Delta_{3}\;\Delta_{2}}\right]\!(\,q),
\end{eqnarray}
where
\begin{equation}
\label{HH}
H_{\!c,\Delta}\!\left[_{\Delta_{4}\;\Delta_{1}}^{\Delta_{3}\;\Delta_{2}}\right]\!(\,q)
=  1 + \sum_{n=1}^\infty
(16q)^{\;n} H^{\,n}_{\!c,\Delta}\!\left[_{\Delta_{4}\;\Delta_{1}}^{\Delta_{3}\;\Delta_{2}}\right].
\end{equation}
The coefficients in (\ref{HH})
are uniquely determined by the recursion relation:
\begin{eqnarray}
\label{recrelHH}
H^{\,n}_{\!c,\,\Delta}\!\left[_{\Delta_{4}\;\Delta_{1}}^{\Delta_{3}\;\Delta_{2}}\right]
&=&
\sum_{\begin{array}{c}\scriptstyle r\geq 1\; s\geq 1\\[-3pt]\scriptstyle n\,\geq \,rs \,\geq 1 \end{array}}
\!\!\!\!\!
\frac{
R^{\,rs}_c\!\left[_{\Delta_{4}\;\Delta_{1}}^{\Delta_{3}\;\Delta_{2}}\right]}
{\Delta-\Delta_{rs}(c)}
\;H^{\,n-rs}_{\!c,\,\Delta_{rs}(c) +rs}\!
\left[_{\Delta_{4}\;\Delta_{1}}^{\Delta_{3}\;\Delta_{2}}\right],
\hskip 1cm
n>0,
\end{eqnarray}
where for the central charge parameterized as $c= 1+6\left(b+{1\over b}\right)^2:$
\begin{eqnarray*}
\Delta_{rs}(c )=
    &=&{1-r^2\over 4}b^2 +{1-rs\over 2} +{1-s^2\over 4} {1\over b^2},
\end{eqnarray*}
and
$R^{\,rs}_c\!\left[_{\Delta_{4}\;\Delta_{1}}^{\Delta_{3}\;\Delta_{2}}\right]$
are  related to the coefficients
$\tilde
R^{\,rs}_\Delta\!\left[_{\Delta_{4}\;\Delta_{1}}^{\Delta_{3}\;\Delta_{2}}\right]$
and their explicit form is known \cite{Zamolodchikov:ie} (see Appendix).
There are some important advantages of the formulae above. First of all one
gets a series which is supposed to converge on the whole domain of analyticity of the conformal block.
Secondly if we choose the  parametrization (\ref{convenient})  of the central charge
the formulae does not contain square roots which simplifies symbolic calculations a lot.

The representation (\ref{Hrep}) and the asymptotic (\ref{defccb}) imply
 the following representation for the classical
conformal block:
\begin{eqnarray}
\label{Hclassblock}
f_{\delta}\!\left[_{\delta_{4}\;\delta_{1}}^{\delta_{3}\;\delta_{2}}\right]\!(\,x)
&=&
 ({\textstyle {1\over 4}}-\delta_1-\delta_2) \log x
+
 ({\textstyle {1\over 4}}-\delta_1-\delta_3) \log (1-x)\\
 \nonumber
&+&
 ({\textstyle {1\over 4}}-2(\delta_1+\delta_2+\delta_3+\delta_4)) \log
 \left( {\textstyle{2\over\pi}}K(x)\right)\\
 \nonumber
 &+& (\delta -{\textstyle {1\over 4}}) \log 16
 - (\delta -{\textstyle {1\over 4}}) \pi {K(1-x)\over K(x)}+
  h_{\delta}\!\left[_{\delta_{4}\;\delta_{1}}^{\delta_{3}\;\delta_{2}}\right](q),
\end{eqnarray}
where
\begin{equation}
\label{hexp}
 h_{\delta}\!\left[_{\delta_{4}\;\delta_{1}}^{\delta_{3}\;\delta_{2}}\right](q)
 =\sum_{n=1}^\infty
(16q)^{\;n} h^{\,n}_{\delta}\!\left[_{\delta_{4}\;\delta_{1}}^{\delta_{3}\;\delta_{2}}\right].
\end{equation}
The coefficients in (\ref{hexp}) can be calculated
 step by step from the asymptotic (\ref{defccb})
\begin{equation}
\label{hlimit}
\sum_{n=1}^\infty
(16q)^{\;n} h^{\,n}_{\delta}\!\left[_{\delta_{4}\;\delta_{1}}^{\delta_{3}\;\delta_{2}}\right] =
\lim\limits_{Q^2 \to \infty} {1\over Q^2} \log\left(1 + \sum_{n=1}^\infty
(16q)^{\;n} H^{\,n}_{\!c,\Delta}\!\left[_{\Delta_{4}\;\Delta_{1}}^{\Delta_{3}\;\Delta_{2}}\right]
\right).
\end{equation}
Using this formula in the special case of all parabolic external weights we have checked
the asymptotic (\ref{defccb}) up to the terms $q^{16}$. We have also calculated the coefficients
of (\ref{hexp}) up to this order. The first few terms of the series (\ref{hexp})
read:
\begin{eqnarray}
\label{q-expansion}
h(p,q)&\equiv &
h_{{1\over 4} +p^2 }\!\left[\;_{{1\over 4}\;\;\;{1\over 4}}^{{1\over 4}\;\;\;{1\over 4}}\;\right]\!(\,q)
\;=\;\textstyle 1+ \frac{1}{4\left( 1 + p^2 \right) }\,q^2
\\ \nonumber
&+& \textstyle\left(
\frac{-1}{32\,{\left( 1 + p^2 \right) }^3} +
  \frac{3}{128\,{\left( 1 + p^2 \right) }^2} +
  \frac{15}{128\,\left( 1 + p^2 \right) } +
  \frac{81}{128\,\left( 4 + p^2 \right) }\right)q^4
\\ \nonumber
&+&\textstyle
\left(
\frac{1}{96\,{\left( 1 + p^2 \right) }^5} -
  \frac{5}{384\,{\left( 1 + p^2 \right) }^4} -
  \frac{9}{256\,{\left( 1 + p^2 \right) }^3} +
  \frac{59}{2048\,{\left( 1 + p^2 \right) }^2} \right.
  \\ \nonumber
  && +\textstyle
 \left.
  \frac{661}{16384\,\left( 1 + p^2 \right) } -
  \frac{9}{128\,\left( 4 + p^2 \right) } +
  \frac{16875}{16384\,\left( 9 + p^2 \right) }
 \right)q^6
 \\ \nonumber
&+&\textstyle
\left(
\frac{-5}{1024\,{\left( 1 + p^2 \right) }^7} +
  \frac{35}{4096\,{\left( 1 + p^2 \right) }^6} +
  \frac{141}{8192\,{\left( 1 + p^2 \right) }^5}\right.
  \\ \nonumber
&&\textstyle -
  \frac{1725}{65536\,{\left( 1 + p^2 \right) }^4}-
  \frac{1653}{65536\,{\left( 1 + p^2 \right) }^3} +
  \frac{6279}{262144\,{\left( 1 + p^2 \right) }^2}
  \\ \nonumber
&&\textstyle +
  \frac{54913}{2097152\,\left( 1 + p^2 \right) } -
  \frac{6561}{16384\,{\left( 4 + p^2 \right) }^3} +
  \frac{19683}{262144\,{\left( 4 + p^2 \right) }^2}
  \\ \nonumber
&&\textstyle + \left.
  \frac{332239}{1048576\,\left( 4 + p^2 \right) } -
  \frac{50625}{2097152\,\left( 9 + p^2 \right) } +
  \frac{1500625}{1048576\,\left( 16 + p^2 \right) }
\right) q^8+\dots.
\end{eqnarray}

\section{Geodesic length function}
\setcounter{equation}{0}

According to the uniformization theorem the punctured sphere $X$
is conformally equivalent to ${\mathbb H}/G,$ where ${\mathbb H}$
is the upper half plane endowed with the Poincare hyperbolic
metric and $G$ is the Fuchsian group uniquely (up to conjugation
in ${\rm PSL}(2,{\mathbb R})$) determined by $X$ and isomorphic to
its fundamental group.

The group $G$ is generated by $T_i \in {\rm PSL}(2,{\mathbb R}),\ i = 1,\ldots,n.$ It is possible
to chose them in such a way that
\[
T_1T_2\ldots T_n =  I.
\]
If all the punctures correspond to parabolic singularities, then
\[
\left|{\rm Tr}\,T_i\right|  =  2.
\]
For each pair of punctures, say $z_i$ and $z_j,$ there exists a unique closed geodesics
of the hyperbolic metrics on $X,$ separating $z_i$ and $z_j$ from the remaining singularities.
Its length $\ell\left(\gamma_{ij}\right)$ can be determined from the relation
\begin{equation}
\label{lenght:ij}
2 \cosh\frac{\ell\left(\gamma_{ij}\right)}{2}  =  \left|{\rm Tr}\, T_iT_j\,\right|.
\end{equation}
Setting
for the four punctured sphere (by an appropriate global conformal transformation)
$z_1 = 0,\ z_2 = x,\ z_3 = 1$ and $z_4 = \infty$ we have in the notation from the previous sections
\(
\ell\left(\gamma_{12}\right) \equiv \ell\left(\gamma_{s}\right), \
\ell\left(\gamma_{23}\right) \equiv \ell\left(\gamma_{t}\right), \
\ell\left(\gamma_{24}\right) \equiv \ell\left(\gamma_{u}\right).
\)

In the case of locations
 $x = \frac19, \frac12, {\rm e}^{-\frac{\pi i}{3}}$ the group $G$ is explicitly  known \cite{Hempel}
$$
\begin{array}{c|c|c|c|c}
x& T_1 &T_2 &T_3 &T_4 \\[2pt]
\hline&&&&\\[-6pt]
\frac19&
\scriptsize
\left(
\begin{array}{rr}
-1 & 0
\\ 2 & -1
\end{array}
\right)
&\scriptsize
\left(\begin{array}{rr}
2 & -3
\\
3 & -4
\end{array}\right)
&\scriptsize
\left(\begin{array}{rr}
2 & -9
\\
1 & -4
\end{array}\right)
&\scriptsize
\left(\begin{array}{rr}
-1 & -6
\\
0 & -1
\end{array}\right)\\[12pt]
\hline&&&&\\[-6pt]
\frac12
&\scriptsize
\left(\begin{array}{rr}
-1 & 0
\\ 2 & -1
\end{array}\right)
&\scriptsize
\left(\begin{array}{rr}
3 & -4
\\
4 & -5
\end{array}\right)
&\scriptsize
\left(\begin{array}{rr}
3 & -8
\\
2 & -5
\end{array}\right)
&\scriptsize
\left(\begin{array}{rr}
-1 & -4
\\
0 & -1
\end{array}\right)\\[12pt]
\hline&&&&\\[-6pt]
{\rm e}^{-\frac{\pi i}{3}}
&\scriptsize
\left(\begin{array}{rr}
-1 & 0
\\ \frac32 & -1
\end{array}\right)
&\scriptsize
\left(\begin{array}{rr}
5 & -6
\\
6 & -7
\end{array}\right)
&\scriptsize
\left(
\begin{array}{rr}
2 & -6
\\
\frac32 & -4
\end{array}
\right)
&\scriptsize
\left(\begin{array}{rr}
-1 & -8
\\
0 & -1
\end{array}\right)
\end{array} \ .
$$
Taking into account the locations
 obtained from $x= \frac19, \frac12, {\rm e}^{-\frac{\pi i}{3}}$
 by $SL(2,\mathbb{C })$ transformations preserving the set $\{0,1,\infty\}$ one gets
\begin{equation}
\label{known}
\begin{array}{c|cccccc|ccc|cc}
x &\;{1\over 9}&-{1\over 8} &\,{8\over 9}&\;{9\over 8}&-8  & 9   \;&\;{1\over 2}&-1& 2
\;&{\rm e}^{-\frac{\pi i}{3}}&{\rm e}^{+\frac{\pi i}{3}}
\\[2pt]\hline&&&&&&&&&&&\\[-8pt]
\cosh\frac{\ell(\gamma_s)}{2}&    2& 2& 5& 8& 5& 8&  3& 3& 7& {7\over 2}& {7\over 2}
\end{array}
\end{equation}
All the cases listed above concern the  metrics on ${\mathbb H}/G,$ induced by the
hyperbolic Poincare metric on $\mathbb{H}$ with the scalar
curvature $-2$. This corresponds to $\mu=4$ in the relation (\ref{glengthII}) which
in terms of the $s$-channel  saddle point Liouville momentum (\ref{saddle})
reads
\begin{equation}
\label{glengthIII}
\ell(\gamma_s(x)) = 4\pi p_s(x).
\end{equation}
Taking into account the explicit formula (\ref{action:mixed:3}) for the 3-point classical action
one can write the saddle point equation (\ref{saddle}) in the form:
\begin{equation}
\label{saddle:1}
-\pi + 2i\log
{\Gamma(1-2ip)\Gamma^2(\frac12 + ip) \over \Gamma(1+2ip)\Gamma^2(\frac12 - ip)}
= \Re \frac{\partial}{\partial p} f(p,x).
\end{equation}
The precision of a numerical solution to this equation is
in practice determined only by the precision of an approximation
to the classical conformal block $f(p,x)$ at hand.

For the locations inside the unit disc $x={1\over 9},-{1\over 8},{1\over 2}, {8\over 9}$
 one can consider approximation given by
 the first $n$ terms  of the $x$-expansion (\ref{x-expansion}).
 We have done the numerical calculations for $n=0,1,\dots,7$.
The deviations from the exact values measured by the differences
$$
\textstyle \Delta(x)=\cosh\left[\frac{1}{2}\ell(\gamma_s(x))\right]-
\cosh\left[2\pi p_s(x)\right]
$$
are presented in Tab.1. They provide a very good confirmation of the formula
(\ref{glengthIII}).\bigskip

\begin{figure}[h]
{\footnotesize
$$
\begin{array}{r|rl|rl|rl|rl|}
n && \;\;\Delta\left({1\over 9}\right) &&\; \Delta\left(-{1\over 8}\right) && \;\;\Delta\left({1\over 2}\right)
 && \Delta\left({8\over 9}\right)\\
[1pt]\hline&&&&&&&&\\[-9pt]
0 && 1.8\times 10^{-2} &  -\!\!\!\!&1.9\times 10^{-2}  && 2.8\times 10^{-1} & & 1.72\\
[1pt]\hline&&&&&&&&\\[-9pt]
1 && 8.5\times 10^{-4}&  & 9.4\times 10^{-4}   && 7.0\times 10^{-2} & & 1.05 \\
[1pt]\hline&&&&&&&&\\[-9pt]
2 && 5.6\times 10^{-5}& -\!\!\!\!& 6.8\times 10^{-5}  && 2.2\times 10^{-2} & & 0.72 \\
[1pt]\hline&&&&&&&&\\[-9pt]
3 &&4.3\times 10^{-6}&  & 5.8\times 10^{-6}   && 8.1\times 10^{-3} & & 0.53 \\
[1pt]\hline&&&&&&&&\\[-9pt]
4 &&3.6\times 10^{-7}&  -\!\!\!\!& 5.4\times 10^{-7}  && 3.1\times 10^{-3} & & 0.40 \\
[1pt]\hline&&&&&&&&\\[-9pt]
5 &&3.2\times 10^{-8}&  & 5.3\times 10^{-8}   && 1.2\times 10^{-3} & & 0.31 \\
[1pt]\hline&&&&&&&&\\[-9pt]
6 &&2.9\times 10^{-9}&  -\!\!\!\!& 5.4\times 10^{-9}  & & 5.3\times 10^{-5} & & 0.24 \\
[1pt]\hline&&&&&&&&\\[-9pt]
7 &&2.7\times 10^{-10}&  & 5.7\times 10^{-10}  && 2.3\times 10^{-5} && 0.19 \\
[1pt]\hline
\end{array}
$$
\centerline{Tab.1}
}
\end{figure}
A much better precision can be achieved if we use the $q$-expansion of $f(p,x)$ (\ref{q-expansion}).
The results of numerical calculations with the approximations of $f(x,p)$ up to the terms
$q^{2k}, k=0,1,\dots,8$ are presented in Tab.2. The agreement with the conjectured exact formula
(\ref{glengthIII}) is perfect.
\begin{figure}
\footnotesize
$$
\begin{array}{r|rl|rl|rl|rl|}
2k && \Delta\left({1\over 9}\right) && \Delta\left(-{1\over 8}\right) && \Delta\left({1\over 2}\right)
 && \Delta\left({8\over 9}\right)\\
[1pt]\hline&&&&&&&&\\[-9pt]
0 &-\!\!\!\!& 4.0\times 10^{-6} &  -\!\!\!\!&4.0\times 10^{-6}  &-\!\!\!\!& 3.9\times 10^{-4} &-\!\!\!\! &1.1 \times 10^{-2}\\
[1pt]\hline&&&&&&&&\\[-9pt]
2 &-\!\!\!\!& 1.0\times 10^{-10}& -\!\!\!\! &1.0\times 10^{-10}   &-\!\!\!\!& 3.7\times 10^{-7} & -\!\!\!\!& 1.0 \times 10^{-4}\\
[1pt]\hline&&&&&&&&\\[-9pt]
4 &-\!\!\!\!& 1.8\times 10^{-15}& & 1.8\times 10^{-15}  &-\!\!\!\!& 5.4\times 10^{-11} & -\!\!\!\!& 2.3\times 10^{-7}\\
[1pt]\hline&&&&&&&&\\[-9pt]
6 &-\!\!\!\!&1.3\times 10^{-15}&  & 1.8\times 10^{-15}   &-\!\!\!\!& 2.5\times 10^{-13} &-\!\!\!\! & 7.2\times 10^{-9}\\
[1pt]\hline&&&&&&&&\\[-9pt]
8 &-\!\!\!\!&1.3\times 10^{-15}&  & 1.8\times 10^{-15}  && 1.8\times 10^{-15} & & 9.9\times 10^{-11} \\
[1pt]\hline&&&&&&&&\\[-9pt]
10 &-\!\!\!\!&1.3\times 10^{-15}&  &1.8\times 10^{-15}   && 1.3\times 10^{-15} & -\!\!\!\!& 2.0\times 10^{-12} \\
[1pt]\hline&&&&&&&&\\[-9pt]
12 &-\!\!\!\!&1.3\times 10^{-15}&  &1.8\times 10^{-15}  & & 1.3\times 10^{-15} & & 3.0\times 10^{-14} \\
[1pt]\hline&&&&&&&&\\[-9pt]
14 &-\!\!\!\!&1.3\times 10^{-15}& & 1.8\times 10^{-15}  & & 1.3\times 10^{-15} & & 4.4\times 10^{-15} \\
[1pt]\hline&&&&&&&&\\[-9pt]
16 &-\!\!\!\!&1.3\times 10^{-15}&  & 1.8\times 10^{-15}  && 1.3\times 10^{-15} && 1.5\times 10^{-14} \\
[1pt]\hline
\end{array}
$$
\centering{Tab.2}\bigskip
\end{figure}

The advantage of the $q$-expansion is that
it converges rapidly for all locations $x$ on  the complex plane except tiny neighborhoods of $x=0,1,\infty$.
One can thus verify the formula (\ref{glengthIII}) for all the cases listed in
(\ref{known}). Using the $q$-expansion up to the terms $q^{16}$ we get the
deviations from the exact values (\ref{known}) closed
to the precision of our numerical calculations set to be slightly less than 16 decimal digits (Tab.3).
\begin{figure}[h]
\footnotesize
$$
\begin{array}[t]{r|rl}
x& &\;\;\;\Delta(x)
\\[1pt]\hline&\\[-8pt]
{1\over 9}&
-\!\!\!\!&
1.3\times 10^{-15}
\\[1pt]\hline&\\[-8pt]
-{1\over 8}&
&1.8\times 10^{-15}
\\[1pt]\hline&\\[-8pt]
{8\over 9}&
&1.5\times 10^{-14}
\\[1pt]\hline&\\[-8pt]
{9\over 8}&
&1.4\times 10^{-13}
\\[1pt]\hline&\\[-8pt]
-8  &-\!\!\!\!
&8.9\times 10^{-15}
\\[1pt]\hline&\\[-8pt]
9&
&1.4\times 10^{-13}
\\[1pt]
\hline
\end{array}
\;\;\;\;\;\;\;\;
\begin{array}[t]{r|rl}
x& &\;\;\;\Delta(x)
\\[1pt]\hline&\\[-8pt]
{1\over 2}&
&1.3\times 10^{-15}
\\[1pt]\hline&\\[-8pt]
-1&
&4.4\times 10^{-15}
\\[1pt]\hline&\\[-8pt]
2 &
-\!\!\!\!&7.1\times 10^{-15}
\\[1pt]\hline&\\[-8pt]
{\rm e}^{-\frac{\pi i}{3}}&
&2.7\times 10^{-15}
\\[1pt]\hline&\\[-8pt]
{\rm e}^{+\frac{\pi i}{3}}&
&2.7\times 10^{-15}
\\[1pt]
\hline
\end{array}
$$
\centering{Tab.3}\bigskip
\end{figure}

As another verification of the formula (\ref{glengthIII}) one can compare the numerical results
with the analytic bounds for the geodesic length function   for the locations $x$ such that the geodesic
$\gamma_s(x)$ is contained in the unit disc \cite{Hempel}:
\begin{equation}
\label{bounds}
{2\pi^2\over \log \left( 256 |x|^{-1} +136\right)}<\ell_s(x)
<{2\pi^2 \over \log\left(\scriptstyle{16 \sqrt{|1-x|}\over |x|}\right)
-{\pi^2\over\scriptstyle 4\log\left({16 \sqrt{|1-x|}\over |x|}\right)}}
\end{equation}
We present sample  numerical calculations of the geodesic length
function  along the rays
\begin{equation}
\label{rays}
x(r) = r {\rm e}^{i{\pi\over 4}k},\;\;\;\;k=0,1,2,3,4.
\end{equation}
The results for the $q$-expansion up to the term $q^{16}$
 are plotted  on Fig.1 for two different ranges of $r$.

\begin{figure}[ht]
\centering
\includegraphics{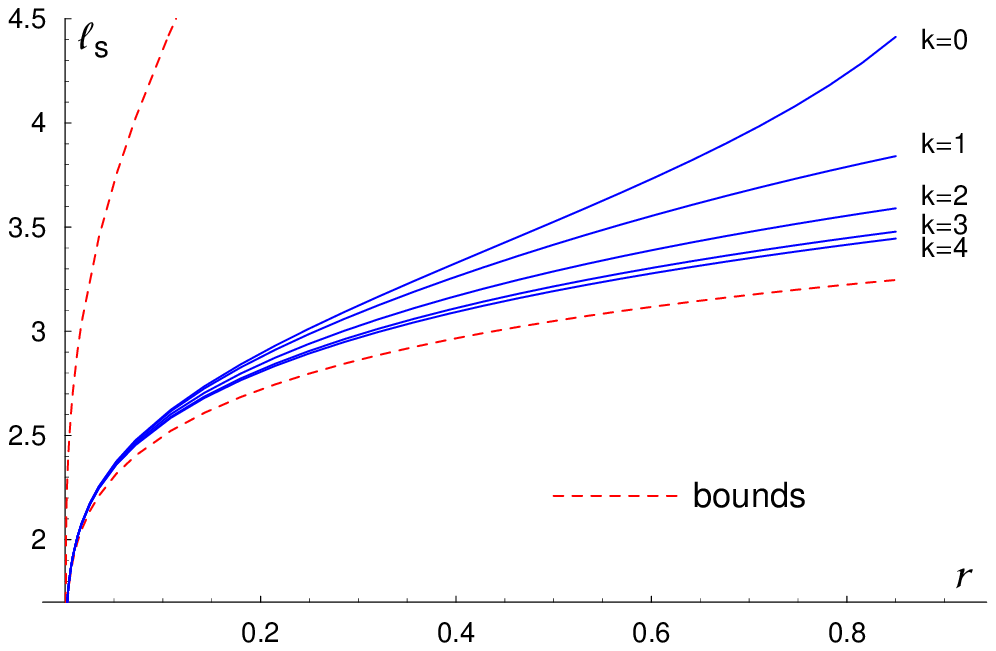}

\includegraphics{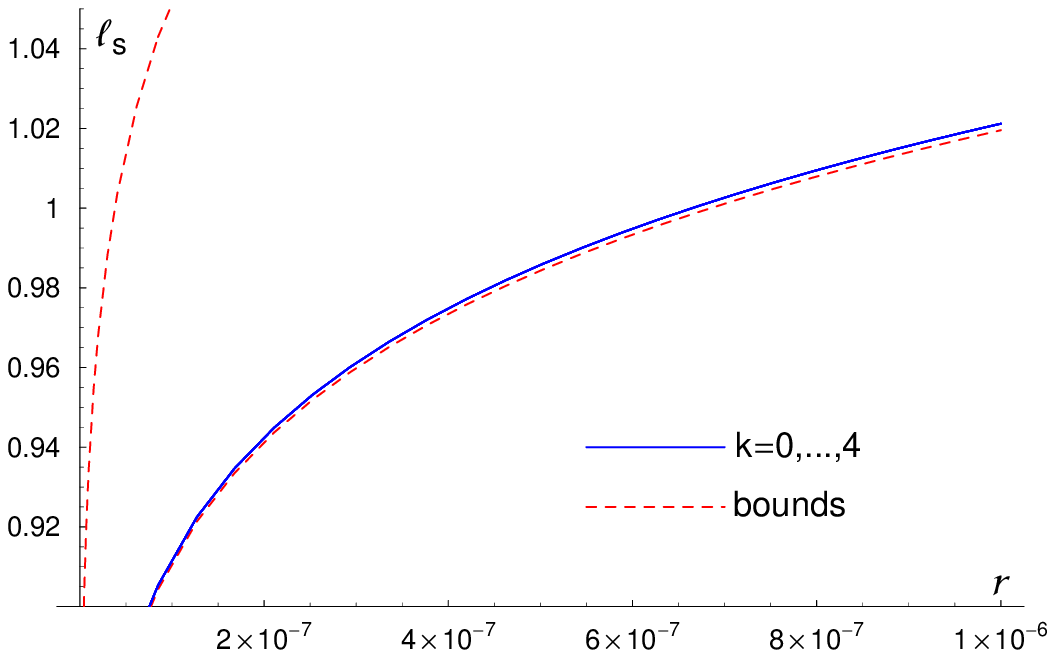}

\footnotesize Fig.1: Numerical calculations of the $s$-channel geodesic length function of the hyperbolic
metric with the constant scalar curvature $R=-2$ along the rays (\ref{rays})
and the analytic bounds (\ref{bounds}).
\end{figure}
Let us close this section with the remark that it is straightforward to work out an
analytic approximation for the saddle point momentum $p_s$ and, consequently,
for the geodesic length $\ell(\gamma_s)$ for small $x.$ Indeed,
with the help of the Lagrange duplication formula
\[
\Gamma(2z) = (2\pi)^{-\frac12} 2^{2z - \frac12}\Gamma(z)\Gamma(z+\frac12)
\]
eq.\ (\ref{saddle:1}) can be rewritten as
\begin{equation}
\label{saddle:2}
-\pi + 16 p \log 2 +2i\log
{\Gamma(1+2ip)\Gamma^2(1 - ip) \over \Gamma(1-2ip)\Gamma^2(1+ip)} =
\Re \frac{\partial}{\partial p} f(p,x).
\end{equation}
Performing a series expansion of the l.h.s.\ according to the formula
\[
\frac{1}{\Gamma(1+z)} =
\exp\left\{\gamma z - \sum\limits_{k=1}^\infty \frac{(-1)^k\zeta(k)}{k} z^k\right\},
\]
where $\gamma$ is the Euler--Marcheroni constant, we can rewrite (\ref{saddle:2}) in the form
\begin{equation}
\label{saddle:3}
-\pi + 16 p \log 2 +
8p^3\sum\limits_{k=1}^\infty\frac{(-1)^{k-1}(1-4^k)\zeta(2k+1)}{2k+1}p^{2k-2} =
\Re \frac{\partial}{\partial p} f(p,x).
\end{equation}
Using the explicit form of $f(p,x)$ one can express a solution of (\ref{saddle:3}) in the
form of an expansion in $\Re\,x$ and $1\over\log x\bar x.$ For instance, keeping in (\ref{saddle:3})
terms up to $p^3$ and $x^3$ we get
\begin{eqnarray*}
p_s & = &
\frac{\pi}{-\log x\bar x + 16\log 2 -\Re\,x - \frac{207}{512}\Re\,x^2 - \frac{205}{1536}\Re\,x^3}
+ \\
&&
\frac{8\zeta(3) +\frac{1}{256}\left(\Re\,x^2 + \Re\,x^3\right)}
{-\log x\bar x + 16\log 2 -\Re\,x - \frac{207}{512}\Re\,x^2 - \frac{205}{1536}\Re\,x^3}\ p_s^3
+{\cal O}\left(\left(\frac{1}{\log x\bar x}\right)^6\!\!,\ \frac{\Re\,x^4}{\log x\bar x}\right)
\\
& = &
\frac{\pi}{-\log x\bar x + 16\log 2 -\Re\,x - \frac{207}{512}\Re\,x^2 - \frac{205}{1536}\Re\,x^3}
+ \\
&&
\frac{8\pi^3\zeta(3) +\frac{\pi^3}{256}\left(\Re\,x^2 + \Re\,x^3\right)}
{\left(-\log x\bar x + 16\log 2 -\Re\,x - \frac{207}{512}\Re\,x^2 - \frac{205}{1536}\Re\,x^3\right)^4}
+{\cal O}\left(\left(\frac{1}{\log x\bar x}\right)^6\!\!,\ \frac{\Re\,x^4}{\log x\bar x}\right).
\end{eqnarray*}

\section{Classical bootstrap}

The $q$-expansion  yields an extremely efficient method of numerical calculation
of the classical conformal block and the saddle point momenta for all locations $x\neq 0,1, \infty$.
The precision of the $s$-channel calculations certainly worsens near $x=1, x=\infty$ singularities.
Still the range of rapid convergence of the $q$-expansion is huge enough to provide a reasonable
testing ground for the classical bootstrap equations (\ref{clasboot}).
For instance in the region $|x|< 4.6\times 10^5$ with the small disc $|1-x|<0.0003$ around $x=1$
removed one has $|q(x)|<{1\over 2}$.

In the present paper we are interested in checking the classical bootstrap equations in the simplest case
of four parabolic singularities.
The $s$-channel factorization of the 4-point classical action (\ref{clasfact}) yields the expression
\begin{eqnarray*}
S_4(x)&\equiv& \textstyle S^{\rm (cl)}\left({1\over 4},{1\over 4},{1\over 4},{1\over 4};x\right) \\
&=& 2S_3(p_s(x)) - 2 \Re f(p_s(x),x)
\end{eqnarray*}
where $f(p,x)$ is the
classical conformal block (\ref{x-expansion}),
$p_s(x)$ is the saddle point momentum in
the $s$-channel given by (\ref{saddle:1}),
and $S_3(p)$ is the 3-point classical action
(\ref{action:mixed:3}) for two parabolic and one hyperbolic weight\footnote{
In order to simplify the formula  an appropriate constant was chosen in (\ref{action:mixed:3}).}:
\begin{eqnarray*}
S_3(p)&\equiv& \textstyle S^{\rm (cl)}\left({1\over 4},{1\over 4},{1\over 4}+p^2\right) \\
&=&\textstyle 4  F\left({1\over 2}+ ip\right) + H(2ip) + \pi p,
\end{eqnarray*}

The efficiency of numerical methods provides a considerable freedom in testing the classical bootstrap
equations.
As an example we have chosen the calculation of the relative deviation from
these equations measured by the functions:
\begin{eqnarray*}
\Delta_{st}(x)&=&  { S_4(1-x)-S_4(x) \over S_4(x)}\\
\Delta_{su}(x)&=&  { S_4(\textstyle {1\over x})-\log |x|-S_4(x) \over S_4(x)}.
\end{eqnarray*}
The largest deviation should be expected around the locations $x=0,1,\infty$.
The behavior of the functions $\Delta_{st}(x),\, \Delta_{su}(x)$ in these regions is well represented
by their values along the real axis. We present the results for two approximations to the classical conformal
block: up to the $q^{12}$ terms and up to the $q^{16}$ terms in the expansion (\ref{q-expansion}).
The results
 are shown on Fig.2 and Fig.3. They provide an excellent numerical
verification of the classical bootstrap.

\begin{figure}[h]
\centering
\includegraphics[width=200pt]{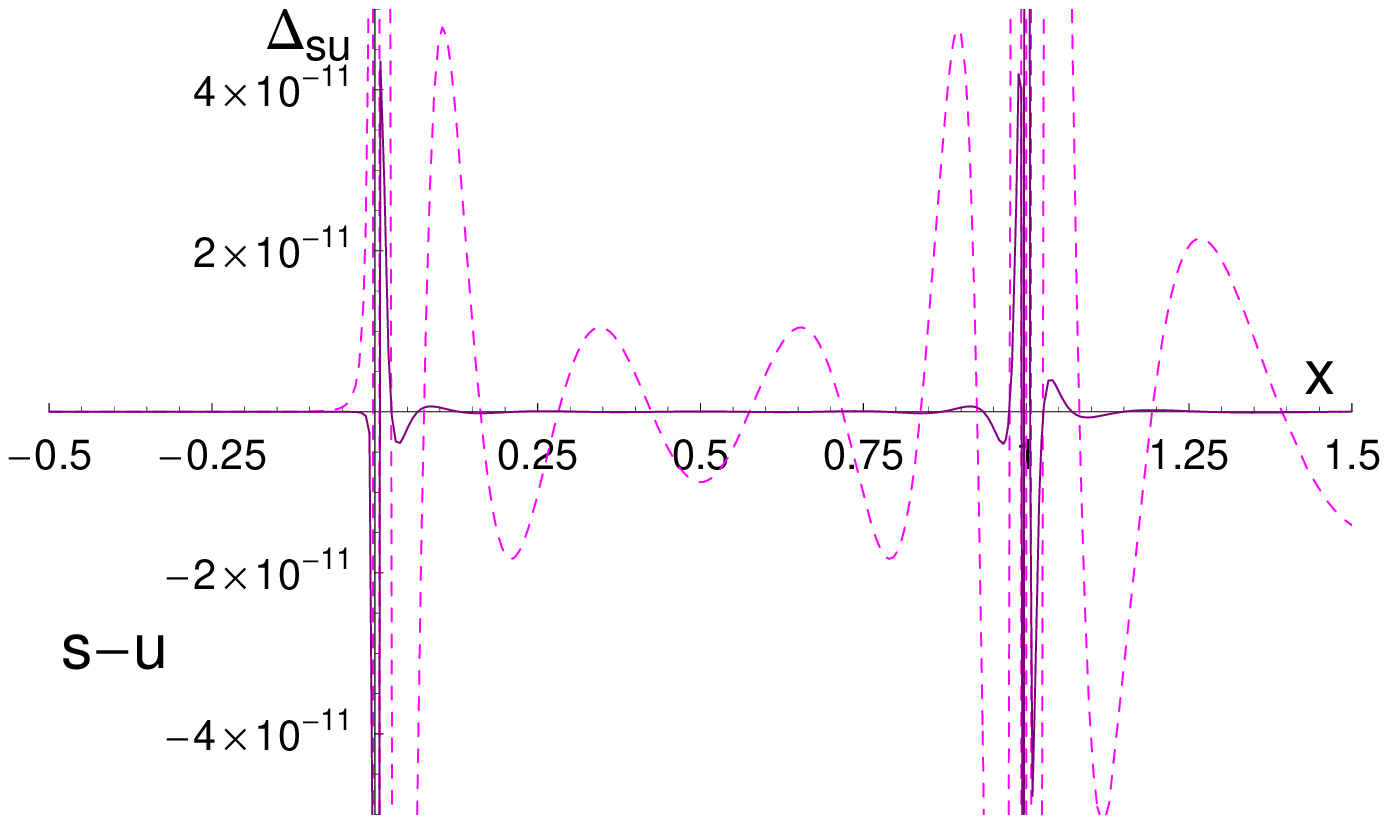}
\includegraphics[width=200pt]{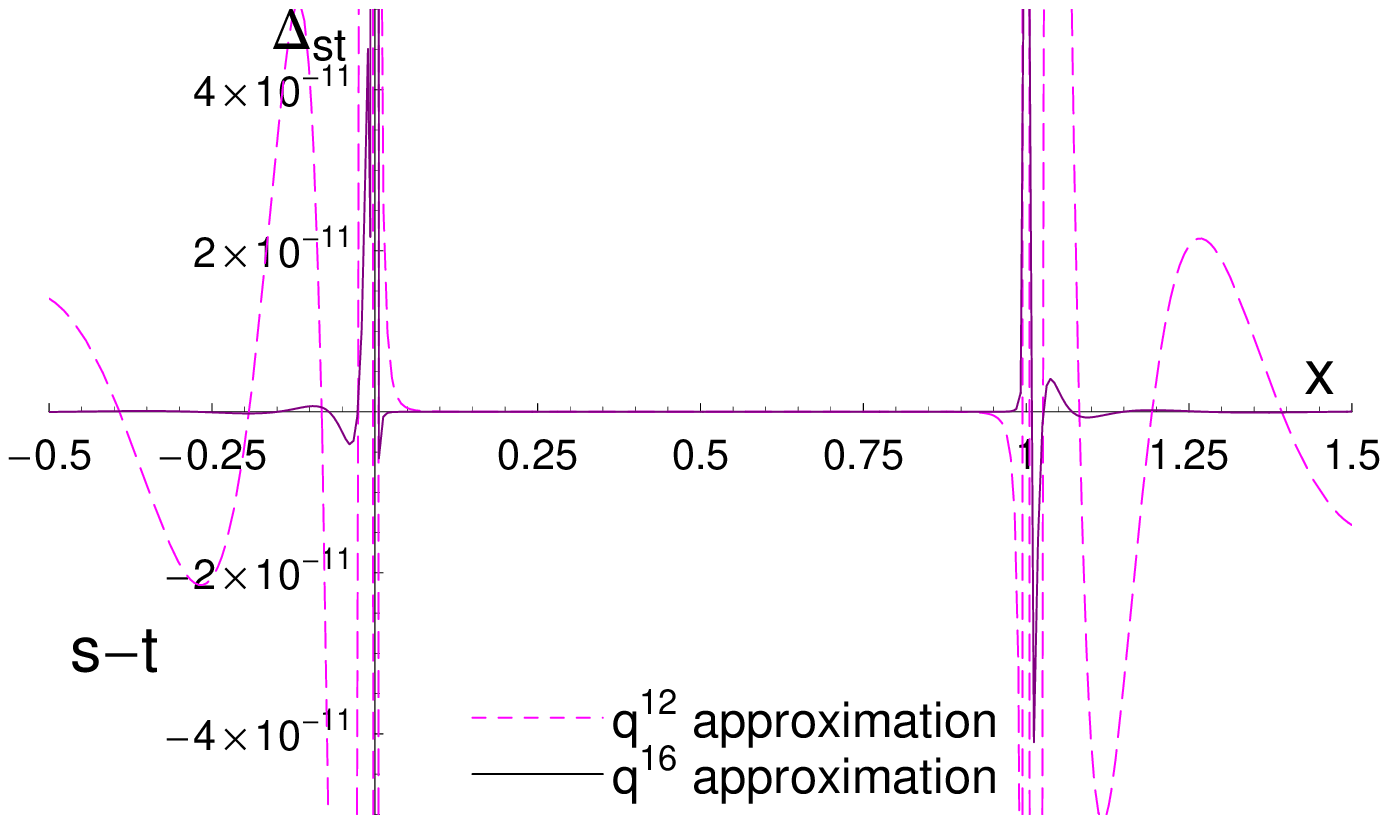}\bigskip

\footnotesize Fig.2: Relative deviation from the classical
bootstrap equations along the real axis near $x=0,1$ singularites.
\end{figure}\bigskip

\begin{figure}[h]
\centering

\includegraphics[width=200pt]{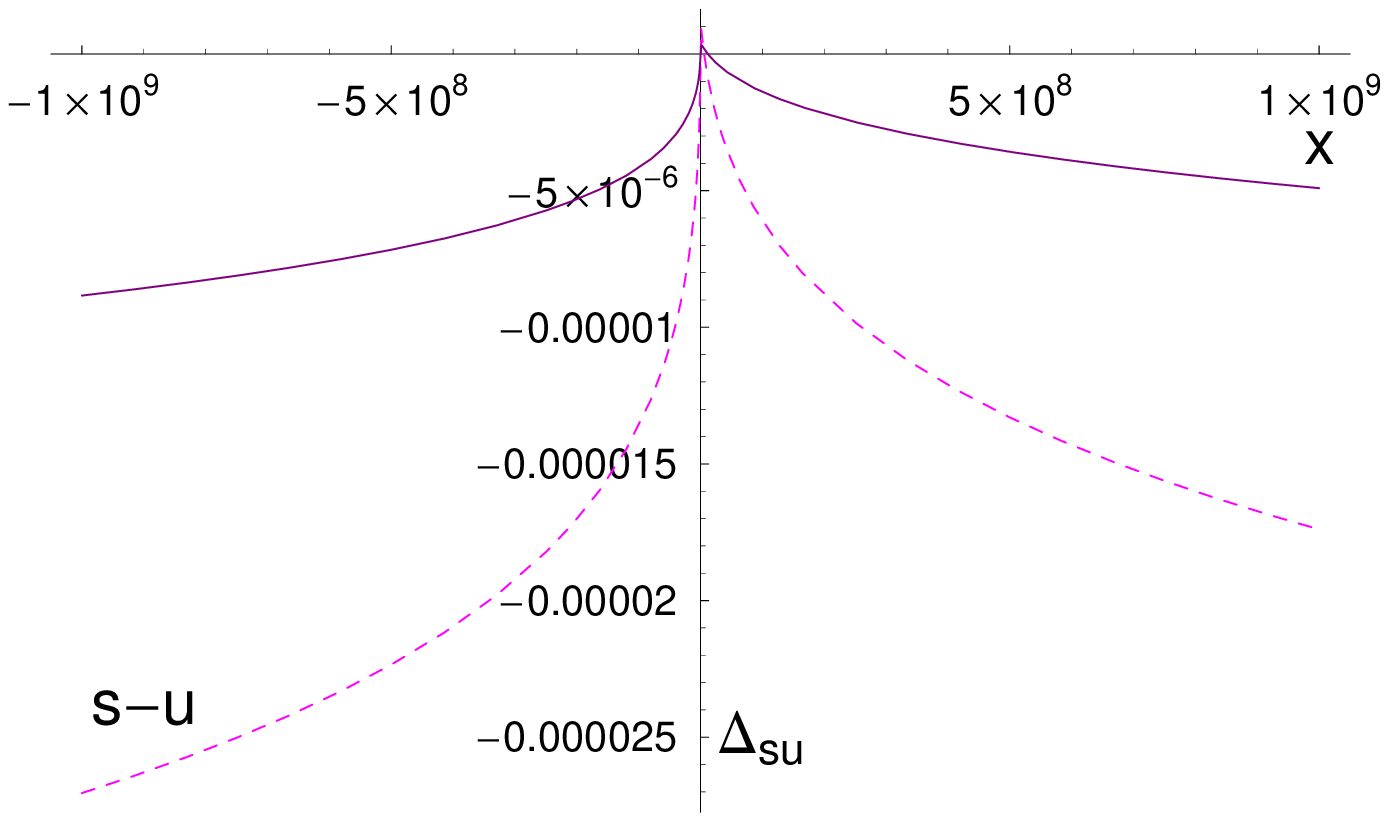}
\includegraphics[width=200pt]{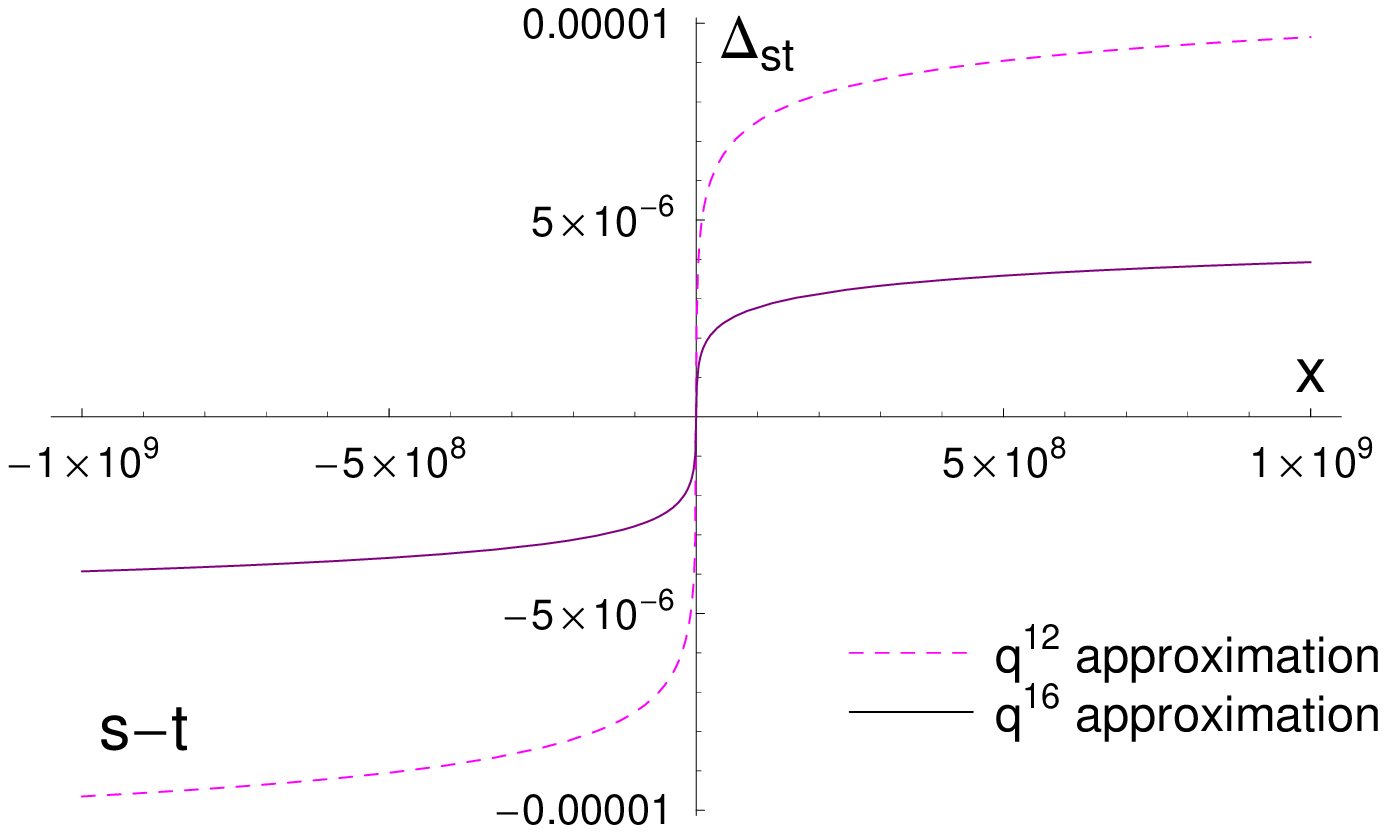}\bigskip

\footnotesize Fig.3: Relative deviation from the classical
bootstrap equations along the real axis near $x=\infty$
singularity.
\end{figure}

\section{Conclusions}

\begin{figure}[p]
\centering
\includegraphics[width=250pt]{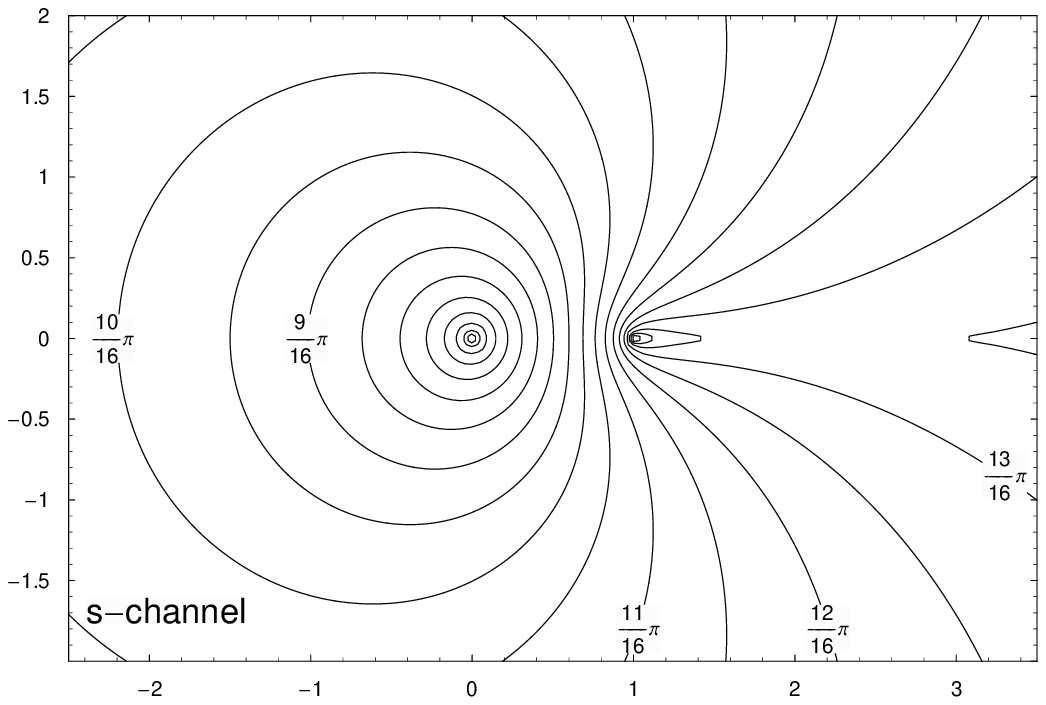}

\includegraphics[width=250pt]{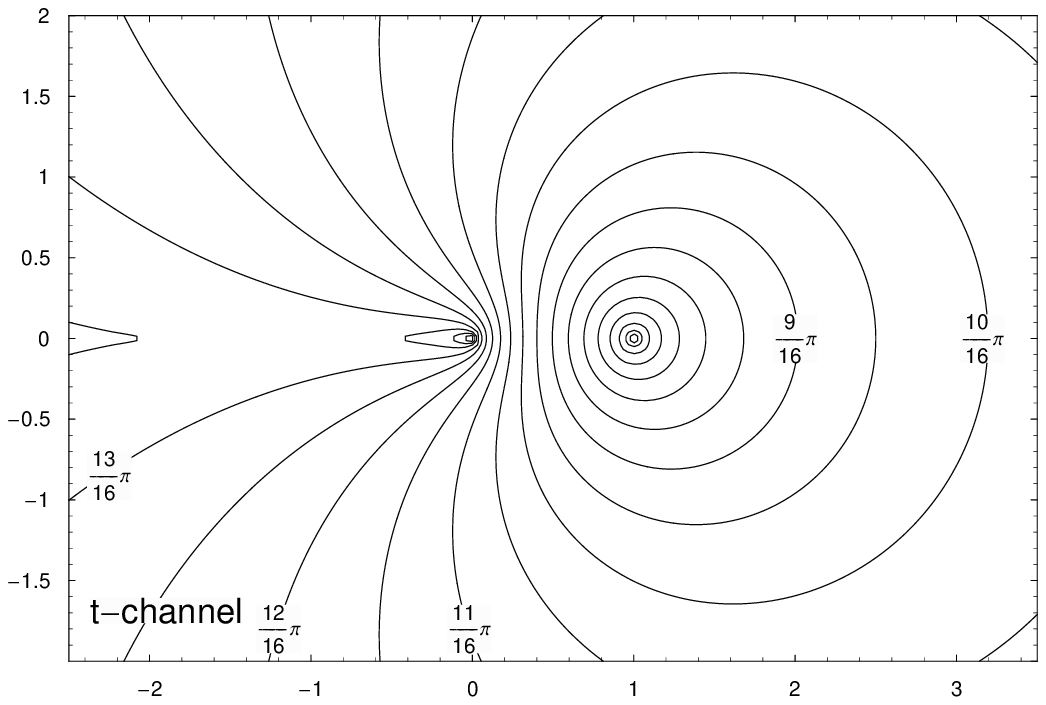}

\includegraphics[width=250pt]{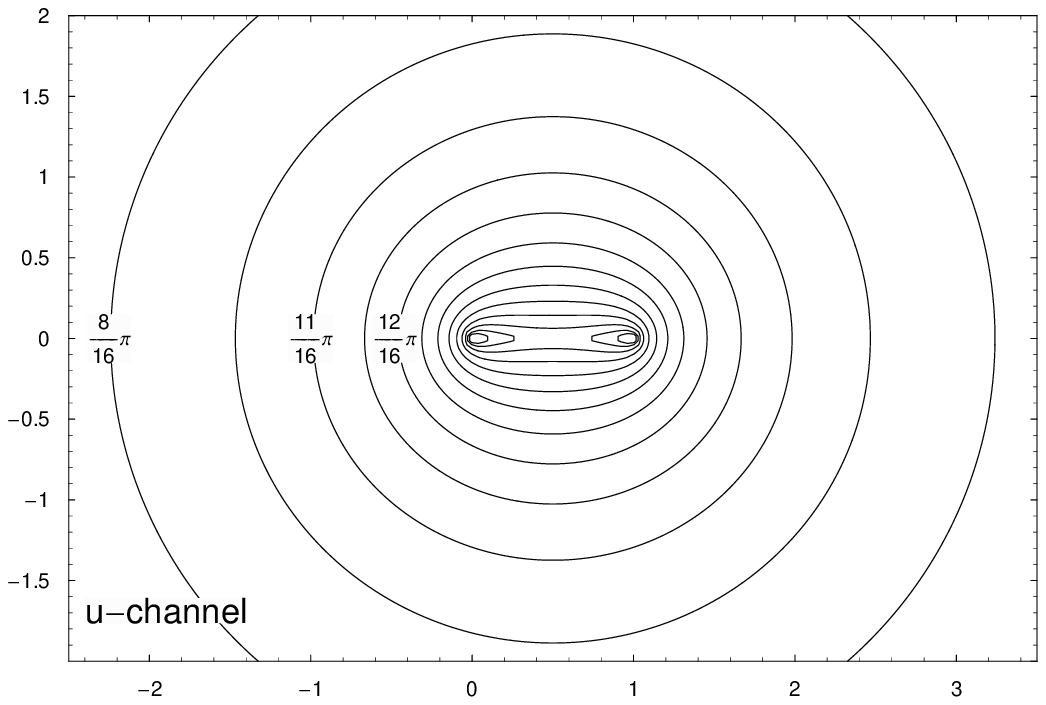}

\footnotesize Fig.4: Geodesic length functions in different channels for the hyperbolic metric\\
with the scalar curvature $R=-1$ (the Gaussian curvature $-{1\over
2}$).
\end{figure}

The numerical tests presented in this paper provide
a convincing evidence that
\begin{itemize}
\item the classical limit of the conformal block exists yielding a consistent definition of
the classical conformal block;
\item the
geodesic length functions of the
hyperbolic geometry on the 4-punctured Riemann sphere can be calculated by the saddle point
equation involving the classical conformal block
and the 3-point classical Liouville action;
\item
the classical Liouville action on the 4-punctured Riemann sphere can be calculated
 in terms of the geodesic length function, the classical conformal block
and the 3-point classical Liouville action.
\end{itemize}
The statements above were derived by heuristic field theoretical
arguments within the path integral representation of the quantum
Liouville theory and should be regarded as well motivated
conjectures. It might be surprising that they are so well supported
by numerical calculations. Still a challenging problem is
to provide rigorous mathematical proofs for them. Any attempt in this direction seem to require
a better understanding of the classical conformal block itself and
in particular a direct way to calculate it. The main motivation
for this line of research is the long standing problem of the
uniformization of the 4-punctured Riemann sphere. Indeed if the
factorization of the classical Liouville action holds and the
classical conformal block and the geodesic length functions are
available
one can calculate the 4-point Liouville action and, by the Polyakov conjecture, the
accessory parameter of the appropriate Fuchsian equation.

The efficiency of numerical calculations based on the
$q$-expansion used in this paper
are completely satisfactory. As an example we present on Fig.4 the contour plots of the
geodesic lengthes in different channels in the case of the
hyperbolic metric with the constant scalar curvature $R=-1,$ i.e.\
the lines of constant value of the length
Frenkel--Nielssen coordinate on the moduli space of the 4-puncture sphere
as a function of the Koba--Nielsen coordinate $x.$
Let us note that also the twist Frenkel--Nielssen coordinate can be calculated.
Indeed it was proved in \cite{ZoTa1,ZoTa2} that the second derivative of
the classical Liouville action with respect to the location of the
punctures, $\partial_{z_i}\partial_{\bar z_j} S^{(\rm cl)},$ gives
the Weil-Petersson metric on the moduli space. With the 4-point classical Liouville
action at hand one can calculate the Weil-Petersson metric and the twist
coordinate along each line of constant length.

\section*{Acknowledgments}
The research of L.H. and Z.J. was supported by the Polish State Research Committee (KBN) grant
no. 1 P03B 025 28.

\section*{Appendix}

\renewcommand{\theequation}{A.\arabic{equation}}
\setcounter{equation}{0} The coefficients $\tilde
R^{\,rs}_\Delta\!\left[_{\Delta_{4}\;\Delta_{1}}^{\Delta_{3}\;\Delta_{2}}\right]$
of the $x$-expansion can be written as
\begin{eqnarray*}
\tilde
R^{\,rs}_\Delta\!\left[_{\Delta_{4}\;\Delta_{1}}^{\Delta_{3}\;\Delta_{2}}\right]
&=&\tilde A^{\,rs}_\Delta \tilde
P^{\,rs}_\Delta\!\left[_{\Delta_{4}}^{\Delta_{3}}\right] \tilde
P^{\,rs}_\Delta\!\left[_{\Delta_{1}}^{\Delta_{2}}\right],
\\[10pt]
\tilde A^{\,rs}_\Delta&=&-{\partial c_{rs}(\Delta)\over \partial
\Delta} {A}_{rs}(T_{rs}(\Delta))= {24(T_{rs}(\Delta)^2 -1)\over
1-s^2 + (r^2 -1)T_{rs}(\Delta)^2}{A}_{rs}(T_{rs}(\Delta)),
 \\[10pt]
\tilde P^{\,rs}_\Delta\!\left[_{\Delta_{1}}^{\Delta_{2}}\right]
&=& P_{rs}(T_{rs}(\Delta),\Delta_1+\Delta_2 ,\Delta_1-\Delta_2 ).
\end{eqnarray*}
For arbitrary positive integers $m,n$ the function
$A_{\,mn}(\alpha^2)$ is defined by the relations
\begin{eqnarray}
\label{A_function} A_{\,mn}(\alpha^2)&=& -{1\over 2}
 \left(\prod\limits_{k=1-m}^m \prod\limits_{l=1-n}^n\right)'
{1\over \displaystyle  k\alpha -{l\over \alpha}},
\end{eqnarray}
where the prime on the symbol of products means that the factors
$(k,l)=(0,0)$ and $(k,l)=(m,n)$ must be omitted.
$A_{\,mn}(\alpha^2)$ can be rewritten in the form
\begin{eqnarray*}
A_{\,mn}(\alpha^2)&=&   {1\over  2mn}
\prod\limits_{k=1}^{m-1}\prod\limits_{l=1}^{n-1}
 {\alpha^4\over (l^2 - k^2 \alpha^4)^2}
\prod\limits_{k=1}^{m-1} {1\over k^2( k^2 \alpha^4-n^2)}
 \prod\limits_{l=1}^{n-1}
 {\alpha^4 \over  l^2 (l^2 - m^2 \alpha^4)}.
\end{eqnarray*}
For arbitrary positive integers $m,n$ the function
${P}_{mn}(\alpha^2,\Delta,\delta)$ is defined by the relations
\begin{eqnarray}
\label{P_function}
P_{mn}(\alpha^2,\Delta_1 +\Delta_2,\Delta_1-\Delta_2)&=&\\
&&\hspace{-150pt}\;=\;
\prod\limits_{\begin{array}{c}\scriptstyle p=1-m\\[-3pt]\scriptstyle p+m=1\, {\rm mod}\, 2
\end{array}}^{m-1}
\!\!\!
\prod\limits_{\begin{array}{c}\scriptstyle q=1-n\\[-3pt]\scriptstyle q+n=1\, {\rm mod}\, 2
\end{array}}^{n-1}
\!\!\!\!\!\!\!\!\! \left({\alpha_1 + \alpha_2 - p\alpha +{q\over
\alpha}\over 2}\right) \left({\alpha_1 - \alpha_2 - p\alpha
+{q\over \alpha}\over 2}\right)\nonumber
\end{eqnarray}
where the  variables  $\alpha_i$ are related to $\Delta_i$ via
$
\Delta_i = -{1\over 4} \left(\alpha - {1\over\alpha}\right)^2
+{\alpha_i^2 \over 4}
$.

\noindent
${P}_{mn}(\alpha^2,\Delta,\delta)$ can be expressed in the form
$$
P_{mn}(\alpha,\Delta,\delta)=\prod\limits_{k=1}^4P^k_{mn}(\alpha,\Delta,\delta)
$$
where
\begin{eqnarray*}
P^1_{mn}(\alpha,\Delta,\delta)&=&\!\!\!\!\!\!\!\!\!
\prod\limits_{\begin{array}{c}\scriptstyle m-1\geq p>0 \\[-3pt]\scriptstyle p+m=1\, {\rm mod}\, 2
\end{array}}
\!\!\!
\prod\limits_{\begin{array}{c}\scriptstyle n-1 \geq q>1-n\\[-3pt]
\scriptstyle q+n=1\, {\rm mod}\, 2
\end{array}}
\!\!\!\!\!\!\!\!\!Q^{p,q}(\alpha,\Delta,\delta)Q^{-p,q}(\alpha,\Delta,\delta)
\\[5pt]
P^2_{mn}(\alpha,\Delta,\delta)\!\!&=&\!\!\left\{
\begin{array}{cl}
\!\!\!\!\prod\limits_{\begin{array}{c}\scriptstyle n-1 \geq q>0\\[-3pt]\scriptstyle q+n=1\, {\rm mod}\, 2
\end{array}}
\!\!\!\!\!\!\!\!\!
Q^{0,q}(\alpha,\Delta,\delta)&{\rm if}\;m \;{\rm is}\; {\rm odd} \\
1&{\rm otherwise}
\end{array}
\right.
\\[15pt]
P^3_{mn}(\alpha,\Delta,\delta)\!\!&=&\!\!\left\{
\begin{array}{cl}
\!\!\!\!\prod\limits_{\begin{array}{c}\scriptstyle m-1 \geq p>0\\[-3pt]\scriptstyle p+m=1\, {\rm mod}\, 2
\end{array}}
\!\!\!\!\!\!\!\!\!
Q^{p,0}(\alpha,\Delta,\delta)&{\rm if}\;n \;{\rm is}\; {\rm odd} \\
1&{\rm otherwise}
\end{array}
\right.
\\[15pt]
P^4_{mn}(\alpha,\Delta,\delta)\!\!&=&\!\!\left\{
\begin{array}{cl}
\delta &{\rm if}\;m \;{\rm and}\;n\;{\rm are}\; {\rm odd}  \\
1&{\rm otherwise}
\end{array}
\right.
\end{eqnarray*}
and
\[
Q^{p,q}(\alpha^2,\Delta,\delta)= \left[{1\over 16}\left({q^2\over
\alpha^2} + p^2\alpha^2 - 2pq\right) \left( {q^2-4\over \alpha^2}
+ (p^2-4)\alpha^2 +2(4-pq) - 8 \Delta\right) + \delta^2
    \right].
\]
The coefficients
$R^{\,rs}_c\!\left[_{\Delta_{4}\;\Delta_{1}}^{\Delta_{3}\;\Delta_{2}}\right]$
that appear in the $q$-expansion of the conformal block can be
written as
\begin{eqnarray*}
R^{\,rs}_c\!\left[_{\Delta_{4}\;\Delta_{1}}^{\Delta_{3}\;\Delta_{2}}\right]
&=&
 A^{\,rs}_c P^{\,rs}_c\!\left[_{\Delta_{4}}^{\Delta_{3}}\right]
P^{\,rs}_c\!\left[_{\Delta_{1}}^{\Delta_{2}}\right],\\[6pt]
A^{\,rs}_c&=&A_{\,rs}(-b^2)=A_{\,rs}(b^2),\\[6pt]
P^{\,rs}_c\!\left[_{\Delta_{1}}^{\Delta_{2}}\right] &=&
P_{rs}(-b^2,\Delta_1+\Delta_2 ,\Delta_1-\Delta_2),
\end{eqnarray*}
where the functions $A_{\,rs}$, and $P_{rs}$ are defined by
(\ref{A_function}) and
(\ref{P_function}) respectively.

\end{document}